\documentclass[reprint,superscriptaddress,amsmath,amssymb,aps,prd,twocolumn,floatfix,nofootinbib]{revtex4-1}
\usepackage{amsmath, amssymb}
\usepackage{listings}
\usepackage{graphicx}
\usepackage{dcolumn}
\usepackage{bm}
\usepackage{float,color}
\usepackage{xcolor}
\usepackage[normalem]{ulem}
\usepackage{tabularx}
\usepackage{mathtools} 
\usepackage{multirow}
\usepackage{amssymb}
\usepackage{scrextend}
\usepackage{hyperref}

\begin{document}

\title{
Multiparameter multipolar test of general relativity with gravitational waves
}

\author{Parthapratim Mahapatra}\email{ppmp75@cmi.ac.in}
\affiliation{Chennai Mathematical Institute, Siruseri, 603103, India}
\author{Shilpa Kastha}
\affiliation{Niels Bohr International Academy, Niels Bohr Institute, Blegdamsvej 17, 2100 Copenhagen, Denmark}
\author{Anuradha Gupta}
\affiliation{Department of Physics and Astronomy, The University of Mississippi, University, Mississippi 38677, USA}
\author{B. S. Sathyaprakash}
\affiliation{Institute for Gravitation and the Cosmos, Department of Physics, Penn State University, University Park, Pennsylvania 16802, USA}
\affiliation{Department of Astronomy and Astrophysics, Penn State University, University Park, Pennsylvania 16802, USA}
\affiliation{School of Physics and Astronomy, Cardiff University, Cardiff, CF24 3AA, United Kingdom}
\author{K. G. Arun}
\affiliation{Chennai Mathematical Institute, Siruseri, 603103, India}
\affiliation{Institute for Gravitation and the Cosmos, Department of Physics, Penn State University, University Park, Pennsylvania 16802, USA}
\date{\today}

\begin{abstract}
Amplitude and phase of the gravitational waveform from compact binary systems can be decomposed in terms of their mass- and current-type multipole moments. In a modified theory of gravity, one or more of these multipole moments could deviate from general theory of relativity. In this work, we show that a waveform model that parametrizes the amplitude and phase in terms of the multipole moments of the binary can facilitate a novel multiparameter test of general relativity with exquisite precision. Using a network of next-generation gravitational-wave observatories, \emph{simultaneous} deviation in the leading seven multipoles of a GW190814-like binary can be bounded to within 6\%--40\% depending on the multipole order, while supermassive black hole mergers observed by the Laser Interferometer Space Antenna achieve a bound of 0.3\%--2\%. 
We further argue that bounds from multipoles can be uniquely mapped onto other parametrized tests of general relativity and have the potential to become a downstream analysis from which bounds of other parametric tests of general relativity can be derived. The set of multipole parameters, therefore, provides an excellent basis to carry out precision tests of general relativity.
\end{abstract}

\maketitle
\section{Introduction}
Gravitational waveform from a compact binary coalescence is a nonlinear function of ``radiative mass-" and ``current-type" multipole moments~\cite{Thorne80} and their derivatives with respect to time. The ``adiabatic inspiral" of the binary is well described by the post-Newtonian (PN) approximation to the general theory of relativity (GR) where the mass ratio and the spins of the binary constituents determine which multipoles are excited and what their contributions are to the emitted flux and the phase evolution of the binary. After the leading quadrupole, the mass-octupole is the next dominant contribution to the phase. As the binary becomes more asymmetric, the contributions from higher-order multipole moments become significant. Spins of the binary constituents can further enhance the strengths of certain higher-order multipoles, especially the current-type ones.

In a modified theory of gravity, where the compact binary dynamics differs from GR, it is natural to expect that one or more of these radiative multipole moments will deviate from those of GR~\cite{Endlich:2017tqa,Compere:2017wrj,Bernard:2018hta,Julie:2019sab,Shiralilou:2021mfl,Battista:2021rlh,Bernard:2022noq,Julie:2022qux,Diedrichs:2023foj}. Therefore, asking whether the measured multipole moments of compact binaries are consistent with GR predictions is an excellent way to test GR. References~\cite{Kastha:2018bcr,Kastha:2019} first derived a multipolar parametrized gravitational-wave phase, which separately tracks the contribution from different radiative multipole moments within the PN approximation to GR. This is achieved by associating parameters $\mu_l$ and $\epsilon_l$ with the mass- and current-type radiative multipole moments, respectively. 
Here $l=2,3,\ldots$ denote quadrupole, octupole, etc. 
The phenomenological multipole parameters are equal to unity in GR (i.e., $\mu_l^{\rm GR}\equiv 1$ and $\epsilon_l^{\rm GR}\equiv 1$), by definition.
By introducing deviations to these multipole coefficients, denoted as $\delta \mu_l$ and $\delta \epsilon_l$ (i.e., $\mu_{l} \equiv 1+\delta \mu_{l}$ and $\epsilon_{l} \equiv 1+\delta \epsilon_{l}$), one can use the gravitational-wave data to obtain bounds on these two sets of parameters.

The radiative multipole moments of compact binaries are nonlinear functionals of the \emph{source} multipole moments (i.e., moments of the stress-energy tensor of the material source and its gravitational fields) and contain time derivatives of the source moments~\cite{BDI95}. These time derivatives of the source multipole moments are evaluated using the equation of motion of the compact binary. Therefore, in the gravitational-wave generation formalism the radiative multipole moments of compact binaries also carry information about the conservative dynamics of the binary. Hence, the parameters $\delta \mu_{l}$ and $\delta \epsilon_{l}$ are sensitive to deviations from GR in both the dissipative and the conservative sectors of the compact binary dynamics. However, one can use the parametrization introduced in Eq.~(3.2) of Ref.~\cite{Kastha:2019} to track explicitly different PN pieces in the conserved orbital energy.

The most general test of GR one can perform, within this framework, is the one where all the $\delta \mu_l$ and $\delta \epsilon_l$ are simultaneously measured, which is often referred to as a ``multiparameter test" (multiparameter tests have been discussed in the context of PN phase expansion in Refs.~\cite{AIQS06a,Gupta:2020lxa, Datta:2020vcj,Saleem:2021nsb}).
We explore the possibility of simultaneously estimating the \emph{leading seven} multipole parameters (i.e., the leading four mass-type and the leading three current-type moments) with the present and next-generation gravitational-wave detectors.
This generalizes the single-parameter projections reported in Refs.~\cite{Kastha:2018bcr,Kastha:2019} and complements the consistency tests proposed in Refs.~\cite{Dhanpal:2018ufk, Islam:2019dmk} and the results from GW190412 and GW190814 being reported in Refs.~\cite{Capano:2020dix, Puecher:2022sfm}. This work also extends the single-parameter octupolar bounds from GW190412 and GW190814 reported recently in Ref.~\cite{Mahapatra:2023hqq}.

The crucial ingredient in this work is the introduction of new parametrized multipolar amplitudes up to 2PN order recently computed in a companion paper~\cite{Mahapatra:2023ydi}, which enables us to use the multipolar information in {\bf both} the amplitude and the phase to derive the bounds on the multipole parameters. Unlike the parametrizations that look for deviations either in 
phase~\cite{BSat94,BSat95,AIQS06a, AIQS06b,YunesPretorius09, MAIS10,Li:2011cg,TIGER,Mehta:2022pcn,TGR-GW150914,TGR-GWTC-1,TGR-GWTC-2,TGR-GWTC-3} or in amplitude~\cite{Islam:2019dmk,Puecher:2022sfm} of gravitational waveform independently, the multipolar parametrization has the advantage that the number of independent parameters is {\bf smaller}, the same as the number of multipole parameters that appear in the {\bf amplitude} and {\bf phase}.

What makes the multiparameter tests very difficult to perform is the strong degeneracies introduced by the simultaneous inclusion of more phenomenological deformation parameters. Multiband gravitational-wave observations~\cite{Gupta:2020lxa,Datta:2020vcj} and principal component analysis~\cite{AP12,Saleem:2021nsb,Datta:2022izc,Datta:2023muk} have been argued to be two different approaches to carry out multiparameter tests of GR in terms of deformations introduced directly in the PN expansion coefficients of the signals's phase evolution. Here, we investigate the use of multipole parameters, as opposed to the usual deformation parameters in the signal's phase, to carry out multiparameter tests of GR.
Apart from being a more downstream parameter set, {\bf orthogonality} of the multipole parameters may help in lifting the above-mentioned degeneracies.

In this work, we show that the multipolar framework is a viable route to carry out a very generic multiparameter test of GR. We further argue that the bounds on $\delta \mu_{l}$ and $\delta \epsilon_{l}$ can be mapped to other parametrized tests of GR. Therefore, this new class of tests may be thought of as an ``all-in-one" test of GR, which may be mapped to any parametrized test of interest. We explicitly demonstrate this mapping in the context of parametrized tests of PN phasing, which is currently employed on the gravitational-wave data and used to obtain constraints on specific modified theories of gravity~\cite{YYP2016}.

The remainder of the paper is organized as follows. In Sec.~\ref{sec:multipolar-waveform}, we briefly describe the parametrized multipolar waveform model. In Sec.~\ref{sec:PE}, we briefly explain the parameter estimation scheme used in our analysis. We discuss our results in Sec.~\ref{sec:Results}. Our conclusions are presented in Sec.~\ref{sec:cncl}.

\section{Waveform model}\label{sec:multipolar-waveform}
We use the frequency-domain amplitude-corrected multipolar waveform for spinning, nonprecessing compact binaries recently reported in Ref.~\cite{Mahapatra:2023ydi}. This waveform model is 3.5PN accurate in the phase and 2PN accurate in the amplitude (i.e., includes the contributions from the first six harmonics). The amplitude-corrected multipolar polarizations in the frequency domain up to 2PN schematically reads~\cite{VanDenBroeck:2006ar, AISS07,ABFO08} 
\begin{equation}
\begin{split}\label{eq:waveform}
    \Tilde{h}_{+,\times} (f) = \frac{G^{2}M^{2}}{c^{5}D_{L}}\sqrt{\frac{5 \, \pi \, \nu}{48}} \sum_{n=0}^{4} \sum_{k=1}^{6} V_{k}^{n-7/2} \, H_{+,\times}^{(k,n)}\; \\
    \times \, e^{i \big( k \Psi_{\rm SPA}(f/k)-\pi/4\big)} \, .
\end{split}    
\end{equation}
Here $M$, $\nu$ ($= \tfrac{q}{(1+q)^{2}}$ with $q$ being the ratio between the primary and secondary mass), and $D_{L}$ denote the redshifted total mass, symmetric mass ratio, and the luminosity distance of the source, respectively. The indices $n$ and $k$ indicate the $\tfrac{n}{2}$th PN order and harmonics of the orbital phase, respectively. The parameter $V_{k}= (2 \, \pi \, G \, M \, f/c^{3}\, k)^{1/3}$ is the dimensionless gauge invariant PN parameter for the $k$th harmonic~\cite{VanDenBroeck:2006ar}, $G$ is the gravitational constant, $c$ is the speed of light, and $f$ is the gravitational-wave frequency. The coefficients $H_{+,\times}^{(k,n)}$ denote the amplitude corrections in the frequency-domain polarizations associated with the contribution from $k$th harmonic at $\tfrac{n}{2}$th PN order. These amplitude coefficients are functions of the masses, spins, and orbital inclination angle $\iota$ and, in our parametrization, contain the multipole parameters $\mu_{l}$ and $\epsilon_{l}$. The expressions for all the $H_{+,\times}^{(k,n)}$ can be found in Eqs.~(10) and (11) of Ref.~\cite{Mahapatra:2023ydi}.
Lastly, $\Psi_{\rm SPA}(f)$ represents the frequency-domain parametrized multipolar gravitational-wave phasing for the first harmonic. References~\cite{Kastha:2019,Kastha:2018bcr} obtained the 3.5PN accurate expression of $\Psi_{\rm SPA}(f)$ for nonprecessing, spinning binaries using the stationary phase approximation.
In the spirit of null tests, the multipolar polarizations in Eq.~(\ref{eq:waveform}) are reexpressed in terms of $\{\delta \mu_{l}, \, \delta \epsilon_{l}\}$ with the goal of deducing projected bounds on them from gravitational-wave observations. 

The gravitational-wave strain in the frequency domain measured by a detector $D$ is given by
\begin{equation}\label{eq:detector-response}
\begin{split}
    \Tilde{h}_D(f)=F_{lp}(f;\theta, \, \phi) \Big[\Tilde{h}_{+}(f) \, F_{+}(f;\theta, \, \phi, \, \psi) \\
    + \Tilde{h}_{\times}(f) \, F_{\times}(f;\theta, \, \phi, \, \psi) \Big]\, ,
\end{split}
\end{equation}
where $F_{lp}$ is the location phase
factor of the detector, $F_{+}$ and $F_{\times}$ are the antenna response functions that describe the detector’s sensitivity to the two different polarizations, $\theta$ is the declination angle, $\phi$ is the right ascension, and $\psi$ is the polarization angle (see Sec. III of Ref.~\cite{gwbench} for more details).

Indeed, our inspiral-only waveform model ignores the contributions from the merger and ringdown phases of the compact binary dynamics, the inclusion of which can lead to a considerable increase in the signal-to-noise ratio (SNR). However, as we crucially make use of the multipole structure in PN theory, it is only natural to employ inspiral-only waveforms for a proof-of-concept study like this, provided we restrict ourselves to binaries that are dominated by their inspiral. Finally, for simplicity, we only consider nonprecessing binary configurations in quasicircular orbits. It is likely that precession- and eccentricity-induced modulations may improve the bounds reported, though the magnitude of this needs to be quantified by a dedicated study.

\begin{table}[t]
\centering
    \begin{tabular}{p{0.25\columnwidth}|p{0.70\columnwidth}}
    \hline
    \hline 
    Network name & Detector location (PSD technology)\\
    \hline
    HLA & LIGO Hanford (A$^\sharp$~\cite{T2200287}), LIGO Livingston (A$^\sharp$), LIGO Aundha~\cite{LIGO-India} (A$^\sharp$)\\
    \hline
    40LA & CE \cite{Evans:2021gyd} Washington (CE 40 km), LIGO Livingston (A$^\sharp$), LIGO Aundha (A$^\sharp$)\\
    \hline 
    40LET & CE Washington (CE 40 km), LIGO Livingston (A$^\sharp$), ET Europe~\cite{Hild:2010id,Punturo:2010zz} (ET 10 km xylophone~\cite{Branchesi:2023mws})\\
    \hline
    4020ET & CE Washington (CE 40 km), CE Texas (CE 20 km), ET Europe (ET 10 km xylophone)\\
    \hline \hline
    \end{tabular}
    \caption{A summary of the four networks of ground-based gravitational-wave detectors used in our analysis. The detector location determines the detector antenna patterns and location phase factors, whereas the PSD technology specifies the used power spectral density. Cosmic Explorer, CE; Einstein Telescope, ET. See Ref.~\cite{Gupta2023mpsac} for more details.
    }
    \label{tab:networks}
\end{table}

\section{Parameter estimation}\label{sec:PE}
To compute the statistical errors on various multipole deformation parameters and other relevant binary parameters, we use the semi-analytical Fisher information matrix formalism~\cite{Rao45,Cramer46,CF94,PW95}.
In the high SNR limit, the Fisher information matrix is a computationally inexpensive method to predict the statistical uncertainties (1$\sigma$ error bars) on the parameters of a signal model buried in stationary Gaussian noise. 

For a frequency-domain gravitational-wave signal $\Tilde{h}_D(f)$, described by a set of parameters $\vec{\lambda}$, the Fisher matrix is defined as
\begin{equation}
    \Gamma_{mn} = 2 \, \int_{f_{\rm min}}^{f_{\rm max}} \frac{\Tilde{h}_{D,m}(f)\,  {\Tilde{h}^{\ast}}_{D,n}(f) +  \Tilde{h}^{\ast}_{D,m}(f)\, \Tilde{h}_{D,n}(f)}{S_{h}(f)} df\, , 
\end{equation}
where $S_{h}(f)$ is the one-sided noise power spectral density (PSD) of the detector, and $f_{\rm min}$ and $f_{\rm max}$ are the lower and upper limits of integration. In the above equation, “$\ast$” denotes the operation of complex conjugation, and “,” denotes differentiation with respect to various elements in the parameter set $\vec{\lambda}\equiv\{\lambda^{m}\}$. The $1\sigma$ statistical error in $\lambda^{m}$ is $\sigma_{m} =\sqrt{\Sigma_{mm}}$, where the covariance matrix $\Sigma_{mn} = (\Gamma_{mn})^{-1}$ is the inverse of the Fisher matrix.

To estimate the errors on all multipole deformation parameters simultaneously, we have considered the following parameter space:
\begin{align}\label{eq:parameters}
    \vec{\lambda}= \Bigl\{t_c, \, \phi_c, \, {\rm log} \mathcal{M}_c, \, \nu, \, \chi_{1z},  \, \chi_{2z},  \, {\rm log} D_{L}, \, \cos \iota, \nonumber\\ 
    \cos \theta, \, \phi, \, \psi, \, \{\delta \mu_l, \, \delta \epsilon_l\}\Bigr\} \, ,
\end{align}
where, $t_{c}$ is the time of coalescence, $\phi_{c}$ is the phase at coalescence, $\mathcal{M}_c=M\, \nu^{3/5}$ is the redshifted chirp mass, and $\chi_{1z}$ and $\chi_{2z}$ are the individual spin components along the orbital angular momentum.\footnote{While estimating the statistical errors on $\{\delta \mu_l, \, \delta \epsilon_l\}$ in the LISA band, we have removed $\cos \theta$, $\phi$, and $\psi$ from the parameter space to improve the inversion accuracy of the Fisher matrix. These parameters are mostly irrelevant for our purposes.} 

For the computation of the statistical errors in the various parameters for different binary configurations and networks of ground-based gravitational-wave detectors, we use {\tt \small GWBENCH}~\cite{gwbench}, a publicly available {\small PYTHON}-based package that computes the Fisher matrix and the corresponding covariance matrix for a given gravitational-wave network. The plus and cross polarizations in Eq.~(\ref{eq:waveform}) are added into {\tt \small GWBENCH} for this purpose. We have chosen $f_{\rm min}$ to be 5 Hz and $f_{\rm max}$ to be 6$F_{\rm ISCO}$ Hz for all the ground-based network configurations. Here $F_{\rm ISCO}$ is the redshifted Kerr innermost stable circular orbit (ISCO) frequency~\cite{Bardeen72, Husa:2015iqa, Hofmann:2016yih} and its explicit expression for nonprecessing binaries can be found in Appendix C of Ref.~\cite{Favata:2021vhw}. For the sources observed by the space-based Laser Interferometer Space Antenna (LISA), we have used Eq.~(2.15) of Ref.~\cite{BBW05a} and have taken $f_{\rm low}=10^{-4}$ Hz and 
$T_{\rm obs}=4$ yr to estimate $f_{\rm min}$.
In the LISA band, $f_{\rm max}$ is given by the smaller of $6 F_{\rm ISCO}$ and 0.1 Hz. We have summarized the different networks of ground-based detectors considered here in Table~\ref{tab:networks}. The noise PSDs of various ground-based detectors used here can be found in {\tt \small GWBENCH}~\cite{gwbench}. 
We have adopted the non-sky-averaged noise PSD of LISA reported in Ref.~\cite{Mangiagli:2020rwz} [see Eqs.~(1)--(5) of \cite{Mangiagli:2020rwz}] and ignored its orbital motion in our computation.

If we assume that all of the multipole deviation parameters take the same value for different events in a population, one can compute a joint bound on them by multiplying the corresponding 1D likelihoods. The width of the joint likelihood is given by
\begin{equation}\label{eq:joint-bound}
    \sigma_{a} = \bigg[ \sum_{i=1}^{N} \left(\sigma_{a}^{(i)} \right)^{-2} \bigg]^{-\frac{1}{2}}\, , \hspace{0.4 cm} a \in \{\delta \mu_l, \, \delta \epsilon_l\}
\end{equation}
where $i =1, \ldots, N $ denotes the events considered in the compact binary population.

\begin{figure}
    \centering
    \includegraphics[scale=0.5]{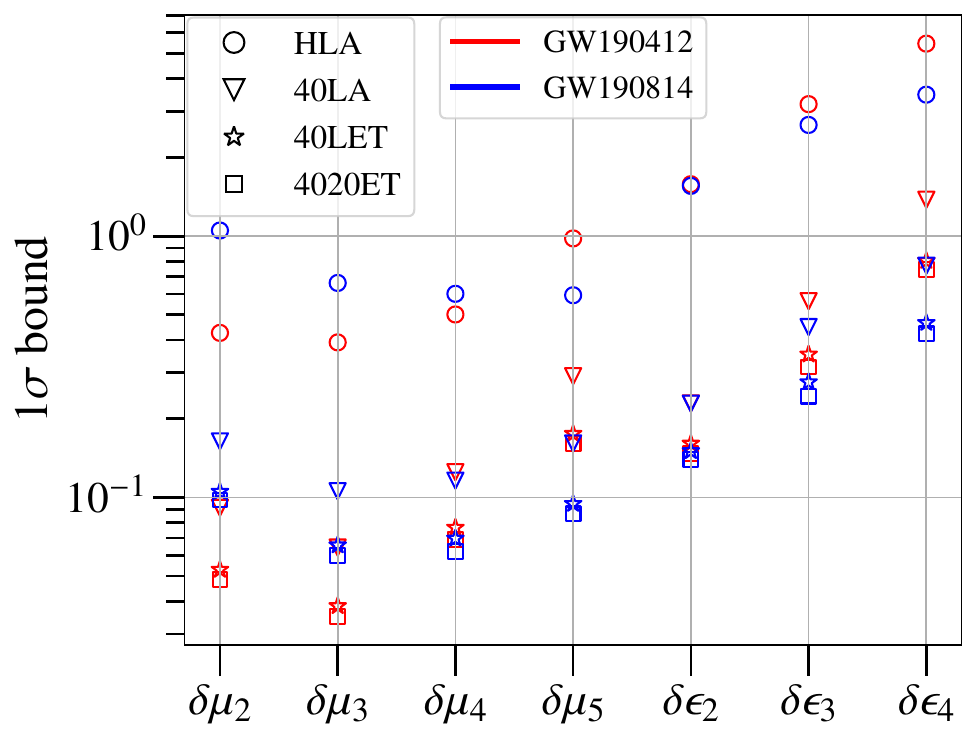}
    \caption{Multiparameter bounds on different multipolar deformation parameters for GW190412- and GW190814-like systems in different networks of future gravitational-wave detectors. Median values from the synthesized population of 100 events is reported (see text for details). Different markers denote different networks considered here.
    }\label{fig:gw190412-gw190814}
\end{figure}



\begin{figure*}
    \centering
    \includegraphics[width=1.0\textwidth]{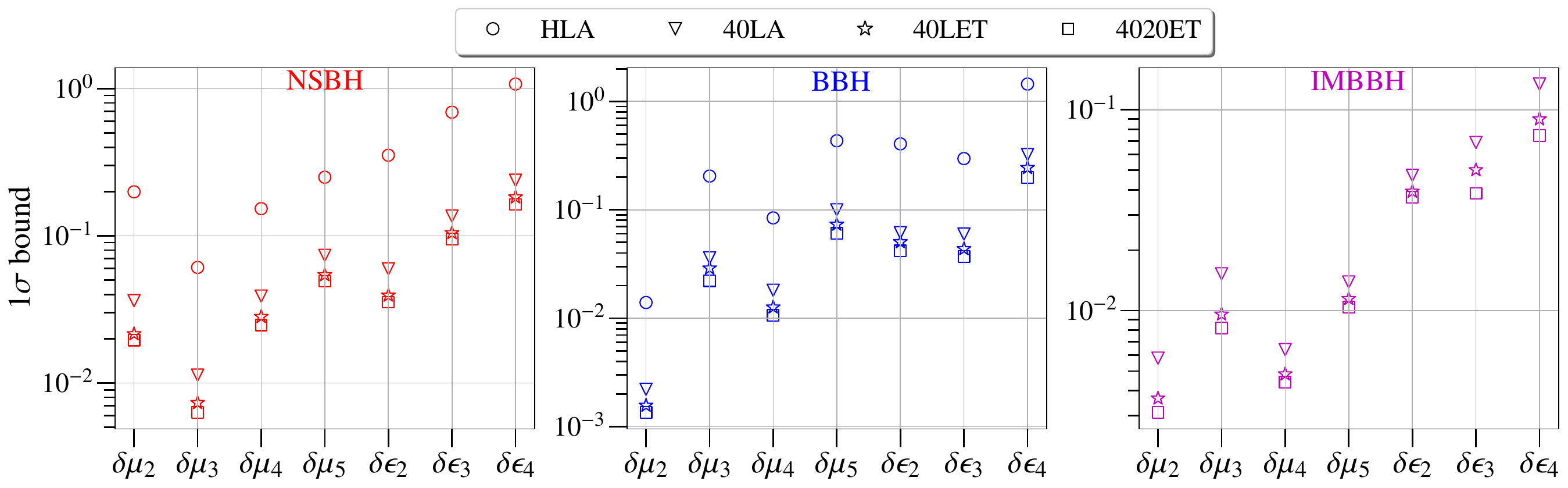}
    \caption{Combined multiparameter bounds on different multipole deformation parameters for three distinct types of compact binary population in different networks of future ground-based gravitational-wave detectors. Population models described in Ref.~\cite{Gupta2023mpsac} are employed, and the loudest 200 events in each category of the source population is combined to obtain the results shown.
    }
    \label{fig:mpsac-pop}
\end{figure*}


\section{Results and discussions}\label{sec:Results}
We start by discussing the projected bounds on the multipole deformation parameters from GW190412-~\cite{GW190412} and GW190814-like systems~\cite{GW190814}, two asymmetric compact binary mergers detected in the third observing run by LIGO/Virgo observatories, in different networks of future ground-based gravitational-wave detectors. As these types of events have been confirmed to exist and extensively studied, they help us to understand the importance of the results. As the observed strengths of the higher-order multipoles depend crucially on the inclination angle $\iota$ and the SNR of the observed gravitational-wave signal depends on the location of the source, we synthesize a population for these two representative systems and use the median value of the resulting distribution to assess the measurement uncertainty in various multipole deformation parameters. Toward this, for each of the systems, we draw 100 samples distributed isotropically over the sphere for the orientation and location of the source. The component masses and spins and the luminosity distances are fixed at the median values reported by Refs.~\cite{GW190412,GW190814,KAGRA:2023pio}. For each sample, we estimate the 1$\sigma$ statistical errors in the seven multipole deformation parameters simultaneously and then compute the median of these 1$\sigma$ errors from the 100 samples. The results for different detector networks are shown in Fig.~\ref{fig:gw190412-gw190814}.

We can measure all seven multipole deformation parameters simultaneously for a GW190814-like system to within $\sim$40\% accuracy in 4020ET, whereas for GW190412-like binaries all multipole deformation parameters can be measured simultaneously to within
$\sim$70\% in 4020ET. Therefore, a single detection of a GW190412- or GW190814-like binary in the next-generation (XG) gravitational-wave detectors will allow us to measure all seven multipole deformation parameters \emph{simultaneously} and hence to perform the \emph{most generic multiparameter} test of gravitational-wave phase and amplitude evolution in GR. It is seen that the mass-type multipole deformation parameters are always estimated better as compared to the current-type multipole deformation parameters. This should be due to the dominance of the mass-type moments over the current-type ones on the dynamics of the binary system. In terms of different detector networks, the 40LET bounds are comparable to those from 4020ET, which suggests that two third-generation detectors already provide very precise bounds and the sensitivity of the third detector does not have a significant impact on the joint bounds.

Next, we consider three different classes of compact binary populations, neutron star--black holes (NSBHs), binary black holes (BBHs), and intermediate mass binary black holes (IMBBHs), reported in Ref.~\cite{Gupta2023mpsac} (see Supplemental Material for details of the population). For each class of the compact binary population, we select 200 loudest events in the respective network of ground-based detectors and calculate the combined bounds on all seven multipole deformation parameters simultaneously using Eq.~(\ref{eq:joint-bound}).

Figure~\ref{fig:mpsac-pop} shows the combined bounds on multipole deformation parameters for these three types of compact binary populations in different networks. We can constrain all the multipole moments simultaneously within an accuracy of $\sim$20\% in the XG era from the NSBH population. The BBH population considered here mostly contains equal-mass binaries, and therefore, they provide the best constraint on $\delta \mu_{2}$. Binaries in the IMBBH population are more massive than the other two populations and are also more asymmetric than the BBH population. As asymmetric massive binaries carry stronger signatures of higher-order multipoles, we obtain the best bounds on higher-order multipole deformation parameters from the IMBBH population---all multipole deformation parameters can be measured simultaneously to within $\sim$8\% in the XG era.  The NSBH population consists mainly of high mass ratio, but less massive, systems than the other two populations. As a result, they provide bounds similar to BBH population on higher-order multipole deformation parameters.


\begin{figure*}
    \centering
    \includegraphics[width=1.0\textwidth]{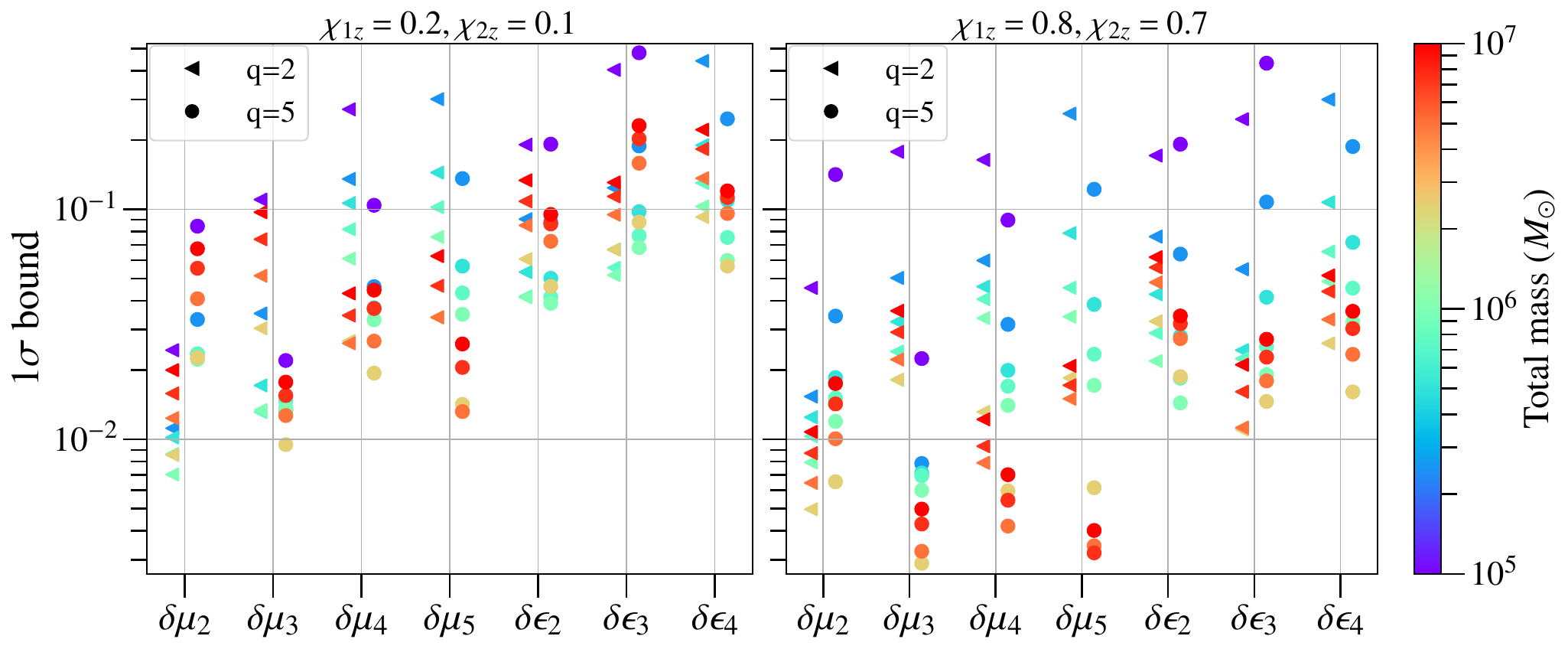}
    \caption{Projected multiparameter constraints on various multipolar deformation parameters for SMBBHs in the LISA band. All the sources are considered to be at a fixed luminosity distance of 3 Gpc. All the angles specifying the binary's orientation and location in the sky are chosen to be $\pi/6$ as a representative angular configuration.
    }
    \label{fig:LISA}
\end{figure*}


The merger rates of supermassive binary black holes (SMBBHs) and their detection rates in LISA are highly uncertain. Here we consider a few representative SMBBH systems and compute the projected error bars on various multipole deformation parameters. We consider merging SMBBHs at a luminosity distance of 3 Gpc with two different choices of spins ($\chi_{1z}=0.2$, $\chi_{2z}=0.1$) and ($\chi_{1z}=0.8$, $\chi_{2z}=0.7$). For each pair of spins, we choose two different mass ratios $2$ and $5$. All the angles (i.e., $\iota$, $\theta$, $\phi$, $\psi$) are set to be $\pi/6$. The 1$\sigma$ errors in all seven deformation parameters in the LISA band for various SMBBH configurations are shown in Fig.~\ref{fig:LISA}. We find that for most of the SMBBH systems considered here, LISA will be able to measure all seven multipole moments simultaneously to within $\sim$10\%.


\begin{figure}
    \centering
    \includegraphics[scale=0.45]{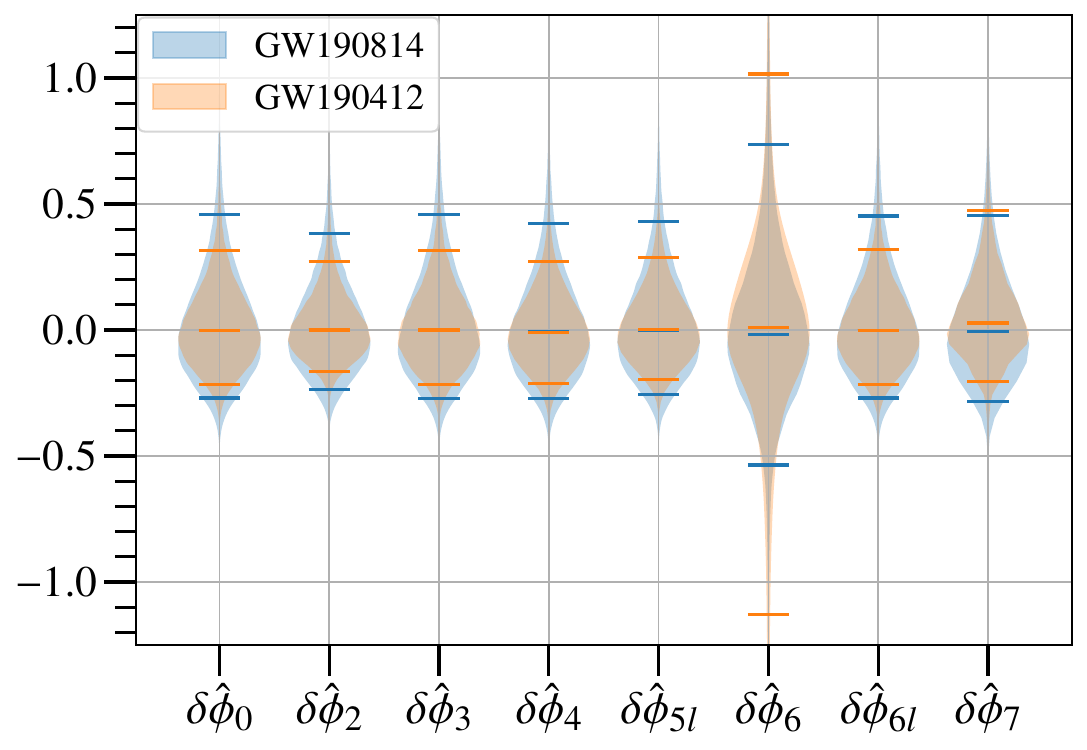}
    \caption{Violin plots for the posterior probability distributions of $\delta \hat{\phi}_{b}$ obtained through the mapping of the multipole deformation bounds for a next-generation detector configuration consisting of two CE and one ET detector ({\tt \small 4020ET}). The horizontal bars indicate the median values and 90\%  credible intervals. }
    \label{fig:MPTGR-TIGER}
\end{figure}

We next discuss how bounds on the PN deformations may be derived from the multipole bounds. In principle, any PN parametrized test of gravitational-wave phase or amplitude can be effectively recast in terms of the multipole parameters. All we need for this is to derive a relation between those phenomenological parameters in the phase or amplitude and $\{\delta \mu_l, \, \delta \epsilon_l\}$. If the parametric form of the phase or amplitude for any test and the contribution of different multipoles to the gravitational-wave phase~\cite{Kastha:2018bcr,Kastha:2019} and amplitude~\cite{Mahapatra:2023ydi} are known, this derivation is straightforward. Here, as a proof-of-principle demonstration, we show how the constraints on $\{\delta \mu_l, \, \delta \epsilon_l\}$ can be mapped onto the different PN deformation parameters $\delta \hat{\phi}_{b}$ in the phase evolution (where $b\in {0, 2, 3, 4, 5l, 6, 6l, 7}$ denotes different PN orders).

Given the gravitational-wave data $d$, we are interested in computing ${\widetilde P}(\delta \hat{\phi}_{b}|d, \, \mathcal{H})$, the posterior probability distribution of $\delta \hat{\phi}_{b}$, for a uniform prior on $\delta \hat{\phi}_{b}$ ($\mathcal{H}$ denotes the hypothesis, which is the parametric model we employ). 
Once we have the  posterior samples for the joint probability distribution ${\widetilde P}(\vec{\lambda}_{I}, \, \vec{\lambda}_{T}|d, \, \mathcal{H})$ for uniform priors on $\vec{\lambda}_{I} \in \{\nu, \, \chi_{1z}, \, \chi_{2z}  \}$ and $\vec{\lambda}_{T} \in \{\delta \mu_{l}, \, \delta \epsilon_{l}\}$,
we can compute the posteriors on $\delta \hat{\phi}_{b}$, $P(\delta \hat{\phi}_{b}|d, \, \mathcal{H})$, using the relation between $\delta \hat{\phi}_{b}$ and $\{\vec{\lambda}_{I}, \, \vec{\lambda}_{T} \}$. As $\delta \hat{\phi}_{b}$ is a unique nonlinear function of $\{\vec{\lambda}_{I}, \, \vec{\lambda}_{T} \}$, a uniform prior on $\{\vec{\lambda}_{I}, \, \vec{\lambda}_{T} \}$ does not translate into  a uniform prior on $\delta \hat{\phi}_{b}$. Therefore, to obtain ${\widetilde P}(\delta \hat{\phi}_{b}|d, \, \mathcal{H})$ we need to reweight the samples of $P(\delta \hat{\phi}_{b}|d, \, \mathcal{H})$ by the samples of $\delta \hat{\phi}_{b}$ derived from the uniform prior on $\{\vec{\lambda}_{I}, \, \vec{\lambda}_{T} \}$, using the relation between $\delta \hat{\phi}_{b}$ and $\{\vec{\lambda}_{I}, \, \vec{\lambda}_{T} \}$. A more detailed discussion about the reweighting procedure is provided in the Supplemental Material.

We consider GW190412- and GW190814-like systems in the {\tt \small 4020ET} network and compute the Fisher matrix $\Gamma_{mn}$ to construct the Gaussian probability distribution function $p(\vec{\lambda}) \propto e^{-\frac{1}{2} \Gamma_{mn}(\lambda^m-\lambda^m_{\rm inj})(\lambda^n-\lambda^n_{\rm inj})}$, where $\lambda^m_{\rm inj}$ are the injected parameter values. We marginalize the distribution $p(\vec{\lambda})$ over parameters other than $\{ \vec{\lambda}_{I}, \,  \vec{\lambda}_{T}\}$ to get ${\widetilde P}(\vec{\lambda}_{I}, \, \vec{\lambda}_{T}\, | \, d, \, \mathcal{H})$. Next, we calculate $P(\delta \hat{\phi}_{b}\, | \, d, \, \mathcal{H})$ using the samples of ${\widetilde P}(\vec{\lambda}_{I}, \, \vec{\lambda}_{T}\, | \, d, \, \mathcal{H})$. To obtain $\widetilde{P}(\delta \hat{\phi}_{b}\, | \, d, \, \mathcal{H})$ that assumes a uniform prior on $\delta \hat{\phi}_{b}$ between $[-10, 10]$, we reweight the distribution $P(\delta \hat{\phi}_{b}\, | \, d, \, \mathcal{H})$ by the distribution of $\delta \hat{\phi}_{b}$ derived from the following prior distributions: $\nu$ is uniform between $[0.045, 0.25]$, $\chi_{1z}$ and $\chi_{2z}$ are uniform between $[-0.99, 0.99]$, and $\vec{\lambda}_{T}$ are uniform between $[-10, 10]$. The posterior distribution $\widetilde{P}(\delta \hat{\phi}_{b}\, | \, d, \, \mathcal{H})$ of different $\delta \hat{\phi}_{b}$ are shown in Fig.~\ref{fig:MPTGR-TIGER}. All the $\delta \hat{\phi}_{b}$ probability distributions are constrained to better than 0.5 at 80\% credibility.

Despite the reweighting employed, the mapped bounds derived here need not match with the regular multiparameter phasing tests using either ground-based or space-based detector alone, where different phasing deformation parameters are treated as independent parameters. This should not be surprising, as the proposed mapping accounts only for the relation between the multipole and the phase deformation parameters and not the correlations these two sets of parameters would have with other binary parameters when the test is performed in the corresponding bases. We have checked that the bounds on the phase deformation parameters derived from the multipole bounds are overall much tighter than those that follow from directly sampling over all of them simultaneously. 

In the case of other parametrized tests of GR that rely on spin-induced multipole moments~\cite{Krishnendu:2017shb, Krishnendu:2018nqa,Krishnendu:2019tjp,Saini:2023gaw},  modified dispersion relations~\cite{Will98,MYW11,Kostelecky:2016kfm,Samajdar:2017mka}, subdominant harmonics~\cite{Islam:2019dmk,Puecher:2022sfm}, etc., the same method will work to derive the corresponding bounds from the multipole ones. In this case, one may visualize the test to be capturing a GR deviation via some \emph{effective} multipolar deformation. A detailed study of these maps and their meanings will be taken up as a follow-up project.

\section{Conclusions and future directions}\label{sec:cncl}
This work serves as a proof-of-concept for the ability of the multipolar framework to carry out a robust multiparameter test of GR with impressive precision, which is necessary to accomplish meaningful constraints on the parameter space of alternate theories of gravity. Moreover, as shown, the bounds from such tests can be uniquely mapped onto the other parametrized tests of GR that rely on amplitude or phase deformations. 

In this work, we have employed the Fisher matrix formalism and a nonprecessing inspiral waveform for parameter estimation. While this paper is meant to illustrate the potential power of the multipolar approach, the results presented here should be revisited using the Bayesian framework with more realistic inspiral-merger-ringdown waveforms. Moreover, the systematic biases induced due to the neglect of well-known effects such as spin precession and eccentricity need to be understood. Hence, the expected constraints that we report here are only indicative of the potential of the multipolar framework.

A natural next step is to construct a parametrized multipolar inspiral-merger-ringdown waveform that includes the effects of spin precession and eccentricity for gravitational-wave data analysis as well as employ state-of-the-art Bayesian parameter inference techniques to demonstrate the feasibility of the method and apply it on a selected subset of gravitational-wave events.

\acknowledgments
The authors thank Elisa Maggio for a critical reading of the manuscript and providing useful comments. We also thank Ish Gupta for a discussion of the compact binary populations used here. P.M. thanks Ssohrab Borhanian for a discussion of {\tt \small GWBENCH}. P.M. also thanks Alok Laddha for valuable discussions. P.M. and K.G.A.~acknowledge the support of the Core Research Grant No. CRG/2021/004565 of the Science and Engineering Research Board of India and a grant from the Infosys Foundation. K.G.A.~acknowledges support from the Department of Science and Technology and Science and Engineering Research Board (SERB) of India via the following grants: Swarnajayanti Fellowship Grant No. DST/SJF/PSA-01/2017-18 and MATRICS grant (Mathematical Research Impact Centric Support) No. MTR/2020/000177. S.K. acknowledges support from the Villum Investigator program supported by the VILLUM Foundation (Grant No. VIL37766) and the DNRF Chair program (Grant No. DNRF162) by the Danish National Research Foundation. This project has received funding from the European Union's Horizon 2020 research and innovation program under the Marie Sklodowska-Curie Grant Agreement No. 101131233. K.G.A. and B.S.S. acknowledge the support of the Indo-U.S. Science and Technology Forum through the Indo-U.S. Centre for Gravitational-Physics and Astronomy, Grant No. IUSSTF/JC-142/2019. We also acknowledge NSF support via NSF Grants No. AST-2205920 and No. PHY-2308887 to A.G. and No. AST-2307147, No. PHY-2012083, No. PHY-2207638, No. PHY-2308886, and No. PHYS-2309064 to B.S.S. This manuscript has the LIGO preprint number P2300424.

The author is grateful for computational resources provided by the LIGO Laboratory and supported by National Science Foundation Grants No. PHY-0757058 and No. PHY-0823459.

This research has made use of data or software obtained from the Gravitational Wave Open Science Center (gwosc.org), a service of the LIGO Scientific Collaboration, the Virgo Collaboration, and KAGRA. This material is based upon work supported by NSF's LIGO Laboratory which is a major facility fully funded by the National Science Foundation, as well as the Science and Technology Facilities Council (STFC) of the United Kingdom, the Max-Planck-Society (MPS), and the State of Niedersachsen/Germany for support of the construction of Advanced LIGO and construction and operation of the GEO600 detector. Additional support for Advanced LIGO was provided by the Australian Research Council. Virgo is funded, through the European Gravitational Observatory (EGO), by the French Centre National de Recherche Scientifique (CNRS), the Italian Istituto Nazionale di Fisica Nucleare (INFN) and the Dutch Nikhef, with contributions by institutions from Belgium, Germany, Greece, Hungary, Ireland, Japan, Monaco, Poland, Portugal, and Spain. KAGRA is supported by Ministry of Education, Culture, Sports, Science and Technology (MEXT), Japan Society for the Promotion of Science (JSPS) in Japan; National Research Foundation (NRF) and Ministry of Science and ICT (MSIT) in Korea; Academia Sinica (AS) and National Science and Technology Council (NSTC) in Taiwan.

\bibliography{ref-list}

\begin{thebibliography}{78}%
\makeatletter
\providecommand \@ifxundefined [1]{%
 \@ifx{#1\undefined}
}%
\providecommand \@ifnum [1]{%
 \ifnum #1\expandafter \@firstoftwo
 \else \expandafter \@secondoftwo
 \fi
}%
\providecommand \@ifx [1]{%
 \ifx #1\expandafter \@firstoftwo
 \else \expandafter \@secondoftwo
 \fi
}%
\providecommand \natexlab [1]{#1}%
\providecommand \enquote  [1]{``#1''}%
\providecommand \bibnamefont  [1]{#1}%
\providecommand \bibfnamefont [1]{#1}%
\providecommand \citenamefont [1]{#1}%
\providecommand \href@noop [0]{\@secondoftwo}%
\providecommand \href [0]{\begingroup \@sanitize@url \@href}%
\providecommand \@href[1]{\@@startlink{#1}\@@href}%
\providecommand \@@href[1]{\endgroup#1\@@endlink}%
\providecommand \@sanitize@url [0]{\catcode `\\12\catcode `\$12\catcode
  `\&12\catcode `\#12\catcode `\^12\catcode `\_12\catcode `\%12\relax}%
\providecommand \@@startlink[1]{}%
\providecommand \@@endlink[0]{}%
\providecommand \url  [0]{\begingroup\@sanitize@url \@url }%
\providecommand \@url [1]{\endgroup\@href {#1}{\urlprefix }}%
\providecommand \urlprefix  [0]{URL }%
\providecommand \Eprint [0]{\href }%
\providecommand \doibase [0]{http://dx.doi.org/}%
\providecommand \selectlanguage [0]{\@gobble}%
\providecommand \bibinfo  [0]{\@secondoftwo}%
\providecommand \bibfield  [0]{\@secondoftwo}%
\providecommand \translation [1]{[#1]}%
\providecommand \BibitemOpen [0]{}%
\providecommand \bibitemStop [0]{}%
\providecommand \bibitemNoStop [0]{.\EOS\space}%
\providecommand \EOS [0]{\spacefactor3000\relax}%
\providecommand \BibitemShut  [1]{\csname bibitem#1\endcsname}%
\let\auto@bib@innerbib\@empty
\bibitem [{\citenamefont {Thorne}(1980)}]{Thorne80}%
  \BibitemOpen
  \bibfield  {author} {\bibinfo {author} {\bibfnamefont {K.~S.}\ \bibnamefont
  {Thorne}},\ }\href {\doibase 10.1103/RevModPhys.52.299} {\bibfield  {journal}
  {\bibinfo  {journal} {Rev. Mod. Phys.}\ }\textbf {\bibinfo {volume} {52}},\
  \bibinfo {pages} {299} (\bibinfo {year} {1980})}\BibitemShut {NoStop}%
\bibitem [{\citenamefont {Endlich}\ \emph {et~al.}(2017)\citenamefont
  {Endlich}, \citenamefont {Gorbenko}, \citenamefont {Huang},\ and\
  \citenamefont {Senatore}}]{Endlich:2017tqa}%
  \BibitemOpen
  \bibfield  {author} {\bibinfo {author} {\bibfnamefont {S.}~\bibnamefont
  {Endlich}}, \bibinfo {author} {\bibfnamefont {V.}~\bibnamefont {Gorbenko}},
  \bibinfo {author} {\bibfnamefont {J.}~\bibnamefont {Huang}}, \ and\ \bibinfo
  {author} {\bibfnamefont {L.}~\bibnamefont {Senatore}},\ }\href {\doibase
  10.1007/JHEP09(2017)122} {\bibfield  {journal} {\bibinfo  {journal} {JHEP}\
  }\textbf {\bibinfo {volume} {09}},\ \bibinfo {pages} {122} (\bibinfo {year}
  {2017})},\ \Eprint {http://arxiv.org/abs/1704.01590} {arXiv:1704.01590
  [gr-qc]} \BibitemShut {NoStop}%
\bibitem [{\citenamefont {Comp\`ere}\ \emph {et~al.}(2018)\citenamefont
  {Comp\`ere}, \citenamefont {Oliveri},\ and\ \citenamefont
  {Seraj}}]{Compere:2017wrj}%
  \BibitemOpen
  \bibfield  {author} {\bibinfo {author} {\bibfnamefont {G.}~\bibnamefont
  {Comp\`ere}}, \bibinfo {author} {\bibfnamefont {R.}~\bibnamefont {Oliveri}},
  \ and\ \bibinfo {author} {\bibfnamefont {A.}~\bibnamefont {Seraj}},\ }\href
  {\doibase 10.1007/JHEP05(2018)054} {\bibfield  {journal} {\bibinfo  {journal}
  {JHEP}\ }\textbf {\bibinfo {volume} {05}},\ \bibinfo {pages} {054} (\bibinfo
  {year} {2018})},\ \Eprint {http://arxiv.org/abs/1711.08806} {arXiv:1711.08806
  [hep-th]} \BibitemShut {NoStop}%
\bibitem [{\citenamefont {Bernard}(2018)}]{Bernard:2018hta}%
  \BibitemOpen
  \bibfield  {author} {\bibinfo {author} {\bibfnamefont {L.}~\bibnamefont
  {Bernard}},\ }\href {\doibase 10.1103/PhysRevD.98.044004} {\bibfield
  {journal} {\bibinfo  {journal} {Phys. Rev. D}\ }\textbf {\bibinfo {volume}
  {98}},\ \bibinfo {pages} {044004} (\bibinfo {year} {2018})},\ \Eprint
  {http://arxiv.org/abs/1802.10201} {arXiv:1802.10201 [gr-qc]} \BibitemShut
  {NoStop}%
\bibitem [{\citenamefont {Juli\'e}\ and\ \citenamefont
  {Berti}(2019)}]{Julie:2019sab}%
  \BibitemOpen
  \bibfield  {author} {\bibinfo {author} {\bibfnamefont {F.-L.}\ \bibnamefont
  {Juli\'e}}\ and\ \bibinfo {author} {\bibfnamefont {E.}~\bibnamefont
  {Berti}},\ }\href {\doibase 10.1103/PhysRevD.100.104061} {\bibfield
  {journal} {\bibinfo  {journal} {Phys. Rev. D}\ }\textbf {\bibinfo {volume}
  {100}},\ \bibinfo {pages} {104061} (\bibinfo {year} {2019})},\ \Eprint
  {http://arxiv.org/abs/1909.05258} {arXiv:1909.05258 [gr-qc]} \BibitemShut
  {NoStop}%
\bibitem [{\citenamefont {Shiralilou}\ \emph {et~al.}(2022)\citenamefont
  {Shiralilou}, \citenamefont {Hinderer}, \citenamefont {Nissanke},
  \citenamefont {Ortiz},\ and\ \citenamefont {Witek}}]{Shiralilou:2021mfl}%
  \BibitemOpen
  \bibfield  {author} {\bibinfo {author} {\bibfnamefont {B.}~\bibnamefont
  {Shiralilou}}, \bibinfo {author} {\bibfnamefont {T.}~\bibnamefont
  {Hinderer}}, \bibinfo {author} {\bibfnamefont {S.~M.}\ \bibnamefont
  {Nissanke}}, \bibinfo {author} {\bibfnamefont {N.}~\bibnamefont {Ortiz}}, \
  and\ \bibinfo {author} {\bibfnamefont {H.}~\bibnamefont {Witek}},\ }\href
  {\doibase 10.1088/1361-6382/ac4196} {\bibfield  {journal} {\bibinfo
  {journal} {Class. Quant. Grav.}\ }\textbf {\bibinfo {volume} {39}},\ \bibinfo
  {pages} {035002} (\bibinfo {year} {2022})},\ \Eprint
  {http://arxiv.org/abs/2105.13972} {arXiv:2105.13972 [gr-qc]} \BibitemShut
  {NoStop}%
\bibitem [{\citenamefont {Battista}\ and\ \citenamefont
  {De~Falco}(2021)}]{Battista:2021rlh}%
  \BibitemOpen
  \bibfield  {author} {\bibinfo {author} {\bibfnamefont {E.}~\bibnamefont
  {Battista}}\ and\ \bibinfo {author} {\bibfnamefont {V.}~\bibnamefont
  {De~Falco}},\ }\href {\doibase 10.1103/PhysRevD.104.084067} {\bibfield
  {journal} {\bibinfo  {journal} {Phys. Rev. D}\ }\textbf {\bibinfo {volume}
  {104}},\ \bibinfo {pages} {084067} (\bibinfo {year} {2021})},\ \Eprint
  {http://arxiv.org/abs/2109.01384} {arXiv:2109.01384 [gr-qc]} \BibitemShut
  {NoStop}%
\bibitem [{\citenamefont {Bernard}\ \emph {et~al.}(2022)\citenamefont
  {Bernard}, \citenamefont {Blanchet},\ and\ \citenamefont
  {Trestini}}]{Bernard:2022noq}%
  \BibitemOpen
  \bibfield  {author} {\bibinfo {author} {\bibfnamefont {L.}~\bibnamefont
  {Bernard}}, \bibinfo {author} {\bibfnamefont {L.}~\bibnamefont {Blanchet}}, \
  and\ \bibinfo {author} {\bibfnamefont {D.}~\bibnamefont {Trestini}},\ }\href
  {\doibase 10.1088/1475-7516/2022/08/008} {\bibfield  {journal} {\bibinfo
  {journal} {JCAP}\ }\textbf {\bibinfo {volume} {08}},\ \bibinfo {pages} {008}
  (\bibinfo {year} {2022})},\ \Eprint {http://arxiv.org/abs/2201.10924}
  {arXiv:2201.10924 [gr-qc]} \BibitemShut {NoStop}%
\bibitem [{\citenamefont {Juli\'e}\ \emph {et~al.}(2023)\citenamefont
  {Juli\'e}, \citenamefont {Baibhav}, \citenamefont {Berti},\ and\
  \citenamefont {Buonanno}}]{Julie:2022qux}%
  \BibitemOpen
  \bibfield  {author} {\bibinfo {author} {\bibfnamefont {F.-L.}\ \bibnamefont
  {Juli\'e}}, \bibinfo {author} {\bibfnamefont {V.}~\bibnamefont {Baibhav}},
  \bibinfo {author} {\bibfnamefont {E.}~\bibnamefont {Berti}}, \ and\ \bibinfo
  {author} {\bibfnamefont {A.}~\bibnamefont {Buonanno}},\ }\href {\doibase
  10.1103/PhysRevD.107.104044} {\bibfield  {journal} {\bibinfo  {journal}
  {Phys. Rev. D}\ }\textbf {\bibinfo {volume} {107}},\ \bibinfo {pages}
  {104044} (\bibinfo {year} {2023})},\ \Eprint
  {http://arxiv.org/abs/2212.13802} {arXiv:2212.13802 [gr-qc]} \BibitemShut
  {NoStop}%
\bibitem [{\citenamefont {Diedrichs}\ \emph {et~al.}(2023)\citenamefont
  {Diedrichs}, \citenamefont {Schmitt},\ and\ \citenamefont
  {Sagunski}}]{Diedrichs:2023foj}%
  \BibitemOpen
  \bibfield  {author} {\bibinfo {author} {\bibfnamefont {R.~F.}\ \bibnamefont
  {Diedrichs}}, \bibinfo {author} {\bibfnamefont {D.}~\bibnamefont {Schmitt}},
  \ and\ \bibinfo {author} {\bibfnamefont {L.}~\bibnamefont {Sagunski}},\
  }\href@noop {} {\  (\bibinfo {year} {2023})},\ \Eprint
  {http://arxiv.org/abs/2311.04274} {arXiv:2311.04274 [gr-qc]} \BibitemShut
  {NoStop}%
\bibitem [{\citenamefont {Kastha}\ \emph {et~al.}(2018)\citenamefont {Kastha},
  \citenamefont {Gupta}, \citenamefont {Arun}, \citenamefont {Sathyaprakash},\
  and\ \citenamefont {Van Den~Broeck}}]{Kastha:2018bcr}%
  \BibitemOpen
  \bibfield  {author} {\bibinfo {author} {\bibfnamefont {S.}~\bibnamefont
  {Kastha}}, \bibinfo {author} {\bibfnamefont {A.}~\bibnamefont {Gupta}},
  \bibinfo {author} {\bibfnamefont {K.~G.}\ \bibnamefont {Arun}}, \bibinfo
  {author} {\bibfnamefont {B.~S.}\ \bibnamefont {Sathyaprakash}}, \ and\
  \bibinfo {author} {\bibfnamefont {C.}~\bibnamefont {Van Den~Broeck}},\ }\href
  {\doibase 10.1103/PhysRevD.98.124033} {\bibfield  {journal} {\bibinfo
  {journal} {Phys. Rev. D}\ }\textbf {\bibinfo {volume} {98}},\ \bibinfo
  {pages} {124033} (\bibinfo {year} {2018})},\ \Eprint
  {http://arxiv.org/abs/1809.10465} {arXiv:1809.10465 [gr-qc]} \BibitemShut
  {NoStop}%
\bibitem [{\citenamefont {Kastha}\ \emph {et~al.}(2019)\citenamefont {Kastha},
  \citenamefont {Gupta}, \citenamefont {Arun}, \citenamefont {Sathyaprakash},\
  and\ \citenamefont {Van Den~Broeck}}]{Kastha:2019}%
  \BibitemOpen
  \bibfield  {author} {\bibinfo {author} {\bibfnamefont {S.}~\bibnamefont
  {Kastha}}, \bibinfo {author} {\bibfnamefont {A.}~\bibnamefont {Gupta}},
  \bibinfo {author} {\bibfnamefont {K.~G.}\ \bibnamefont {Arun}}, \bibinfo
  {author} {\bibfnamefont {B.~S.}\ \bibnamefont {Sathyaprakash}}, \ and\
  \bibinfo {author} {\bibfnamefont {C.}~\bibnamefont {Van Den~Broeck}},\ }\href
  {\doibase 10.1103/PhysRevD.100.044007} {\bibfield  {journal} {\bibinfo
  {journal} {Phys. Rev. D}\ }\textbf {\bibinfo {volume} {100}},\ \bibinfo
  {pages} {044007} (\bibinfo {year} {2019})}\BibitemShut {NoStop}%
\bibitem [{\citenamefont {Blanchet}\ \emph {et~al.}(1995)\citenamefont
  {Blanchet}, \citenamefont {Damour},\ and\ \citenamefont {Iyer}}]{BDI95}%
  \BibitemOpen
  \bibfield  {author} {\bibinfo {author} {\bibfnamefont {L.}~\bibnamefont
  {Blanchet}}, \bibinfo {author} {\bibfnamefont {T.}~\bibnamefont {Damour}}, \
  and\ \bibinfo {author} {\bibfnamefont {B.~R.}\ \bibnamefont {Iyer}},\
  }\href@noop {} {\bibfield  {journal} {\bibinfo  {journal} {Phys. Rev. D}\
  }\textbf {\bibinfo {volume} {51}},\ \bibinfo {pages} {5360} (\bibinfo {year}
  {1995})},\ \Eprint {http://arxiv.org/abs/gr-qc/9501029} {gr-qc/9501029}
  \BibitemShut {NoStop}%
\bibitem [{\citenamefont {Arun}\ \emph
  {et~al.}(2006{\natexlab{a}})\citenamefont {Arun}, \citenamefont {Iyer},
  \citenamefont {Qusailah},\ and\ \citenamefont {Sathyaprakash}}]{AIQS06a}%
  \BibitemOpen
  \bibfield  {author} {\bibinfo {author} {\bibfnamefont {K.~G.}\ \bibnamefont
  {Arun}}, \bibinfo {author} {\bibfnamefont {B.~R.}\ \bibnamefont {Iyer}},
  \bibinfo {author} {\bibfnamefont {M.~S.~S.}\ \bibnamefont {Qusailah}}, \ and\
  \bibinfo {author} {\bibfnamefont {B.~S.}\ \bibnamefont {Sathyaprakash}},\
  }\href@noop {} {\bibfield  {journal} {\bibinfo  {journal} {Class. Quantum
  Grav.}\ }\textbf {\bibinfo {volume} {23}},\ \bibinfo {pages} {L37} (\bibinfo
  {year} {2006}{\natexlab{a}})},\ \Eprint {http://arxiv.org/abs/gr-qc/0604018}
  {gr-qc/0604018} \BibitemShut {NoStop}%
\bibitem [{\citenamefont {Gupta}\ \emph {et~al.}(2020)\citenamefont {Gupta},
  \citenamefont {Datta}, \citenamefont {Kastha}, \citenamefont {Borhanian},
  \citenamefont {Arun},\ and\ \citenamefont {Sathyaprakash}}]{Gupta:2020lxa}%
  \BibitemOpen
  \bibfield  {author} {\bibinfo {author} {\bibfnamefont {A.}~\bibnamefont
  {Gupta}}, \bibinfo {author} {\bibfnamefont {S.}~\bibnamefont {Datta}},
  \bibinfo {author} {\bibfnamefont {S.}~\bibnamefont {Kastha}}, \bibinfo
  {author} {\bibfnamefont {S.}~\bibnamefont {Borhanian}}, \bibinfo {author}
  {\bibfnamefont {K.~G.}\ \bibnamefont {Arun}}, \ and\ \bibinfo {author}
  {\bibfnamefont {B.~S.}\ \bibnamefont {Sathyaprakash}},\ }\href {\doibase
  10.1103/PhysRevLett.125.201101} {\bibfield  {journal} {\bibinfo  {journal}
  {Phys. Rev. Lett.}\ }\textbf {\bibinfo {volume} {125}},\ \bibinfo {pages}
  {201101} (\bibinfo {year} {2020})},\ \Eprint
  {http://arxiv.org/abs/2005.09607} {arXiv:2005.09607 [gr-qc]} \BibitemShut
  {NoStop}%
\bibitem [{\citenamefont {Datta}\ \emph {et~al.}(2021)\citenamefont {Datta},
  \citenamefont {Gupta}, \citenamefont {Kastha}, \citenamefont {Arun},\ and\
  \citenamefont {Sathyaprakash}}]{Datta:2020vcj}%
  \BibitemOpen
  \bibfield  {author} {\bibinfo {author} {\bibfnamefont {S.}~\bibnamefont
  {Datta}}, \bibinfo {author} {\bibfnamefont {A.}~\bibnamefont {Gupta}},
  \bibinfo {author} {\bibfnamefont {S.}~\bibnamefont {Kastha}}, \bibinfo
  {author} {\bibfnamefont {K.~G.}\ \bibnamefont {Arun}}, \ and\ \bibinfo
  {author} {\bibfnamefont {B.~S.}\ \bibnamefont {Sathyaprakash}},\ }\href
  {\doibase 10.1103/PhysRevD.103.024036} {\bibfield  {journal} {\bibinfo
  {journal} {Phys. Rev. D}\ }\textbf {\bibinfo {volume} {103}},\ \bibinfo
  {pages} {024036} (\bibinfo {year} {2021})},\ \Eprint
  {http://arxiv.org/abs/2006.12137} {arXiv:2006.12137 [gr-qc]} \BibitemShut
  {NoStop}%
\bibitem [{\citenamefont {Saleem}\ \emph {et~al.}(2022)\citenamefont {Saleem},
  \citenamefont {Datta}, \citenamefont {Arun},\ and\ \citenamefont
  {Sathyaprakash}}]{Saleem:2021nsb}%
  \BibitemOpen
  \bibfield  {author} {\bibinfo {author} {\bibfnamefont {M.}~\bibnamefont
  {Saleem}}, \bibinfo {author} {\bibfnamefont {S.}~\bibnamefont {Datta}},
  \bibinfo {author} {\bibfnamefont {K.~G.}\ \bibnamefont {Arun}}, \ and\
  \bibinfo {author} {\bibfnamefont {B.~S.}\ \bibnamefont {Sathyaprakash}},\
  }\href {\doibase 10.1103/PhysRevD.105.084062} {\bibfield  {journal} {\bibinfo
   {journal} {Phys. Rev. D}\ }\textbf {\bibinfo {volume} {105}},\ \bibinfo
  {pages} {084062} (\bibinfo {year} {2022})},\ \Eprint
  {http://arxiv.org/abs/2110.10147} {arXiv:2110.10147 [gr-qc]} \BibitemShut
  {NoStop}%
\bibitem [{\citenamefont {Dhanpal}\ \emph {et~al.}(2019)\citenamefont
  {Dhanpal}, \citenamefont {Ghosh}, \citenamefont {Mehta}, \citenamefont
  {Ajith},\ and\ \citenamefont {Sathyaprakash}}]{Dhanpal:2018ufk}%
  \BibitemOpen
  \bibfield  {author} {\bibinfo {author} {\bibfnamefont {S.}~\bibnamefont
  {Dhanpal}}, \bibinfo {author} {\bibfnamefont {A.}~\bibnamefont {Ghosh}},
  \bibinfo {author} {\bibfnamefont {A.~K.}\ \bibnamefont {Mehta}}, \bibinfo
  {author} {\bibfnamefont {P.}~\bibnamefont {Ajith}}, \ and\ \bibinfo {author}
  {\bibfnamefont {B.~S.}\ \bibnamefont {Sathyaprakash}},\ }\href {\doibase
  10.1103/PhysRevD.99.104056} {\bibfield  {journal} {\bibinfo  {journal} {Phys.
  Rev. D}\ }\textbf {\bibinfo {volume} {99}},\ \bibinfo {pages} {104056}
  (\bibinfo {year} {2019})},\ \Eprint {http://arxiv.org/abs/1804.03297}
  {arXiv:1804.03297 [gr-qc]} \BibitemShut {NoStop}%
\bibitem [{\citenamefont {Islam}\ \emph {et~al.}(2020)\citenamefont {Islam},
  \citenamefont {Mehta}, \citenamefont {Ghosh}, \citenamefont {Varma},
  \citenamefont {Ajith},\ and\ \citenamefont {Sathyaprakash}}]{Islam:2019dmk}%
  \BibitemOpen
  \bibfield  {author} {\bibinfo {author} {\bibfnamefont {T.}~\bibnamefont
  {Islam}}, \bibinfo {author} {\bibfnamefont {A.~K.}\ \bibnamefont {Mehta}},
  \bibinfo {author} {\bibfnamefont {A.}~\bibnamefont {Ghosh}}, \bibinfo
  {author} {\bibfnamefont {V.}~\bibnamefont {Varma}}, \bibinfo {author}
  {\bibfnamefont {P.}~\bibnamefont {Ajith}}, \ and\ \bibinfo {author}
  {\bibfnamefont {B.~S.}\ \bibnamefont {Sathyaprakash}},\ }\href {\doibase
  10.1103/PhysRevD.101.024032} {\bibfield  {journal} {\bibinfo  {journal}
  {Phys. Rev. D}\ }\textbf {\bibinfo {volume} {101}},\ \bibinfo {pages}
  {024032} (\bibinfo {year} {2020})},\ \Eprint
  {http://arxiv.org/abs/1910.14259} {arXiv:1910.14259 [gr-qc]} \BibitemShut
  {NoStop}%
\bibitem [{\citenamefont {Capano}\ and\ \citenamefont
  {Nitz}(2020)}]{Capano:2020dix}%
  \BibitemOpen
  \bibfield  {author} {\bibinfo {author} {\bibfnamefont {C.~D.}\ \bibnamefont
  {Capano}}\ and\ \bibinfo {author} {\bibfnamefont {A.~H.}\ \bibnamefont
  {Nitz}},\ }\href {\doibase 10.1103/PhysRevD.102.124070} {\bibfield  {journal}
  {\bibinfo  {journal} {Phys. Rev. D}\ }\textbf {\bibinfo {volume} {102}},\
  \bibinfo {pages} {124070} (\bibinfo {year} {2020})},\ \Eprint
  {http://arxiv.org/abs/2008.02248} {arXiv:2008.02248 [gr-qc]} \BibitemShut
  {NoStop}%
\bibitem [{\citenamefont {Puecher}\ \emph {et~al.}(2022)\citenamefont
  {Puecher}, \citenamefont {Kalaghatgi}, \citenamefont {Roy}, \citenamefont
  {Setyawati}, \citenamefont {Gupta}, \citenamefont {Sathyaprakash},\ and\
  \citenamefont {Van Den~Broeck}}]{Puecher:2022sfm}%
  \BibitemOpen
  \bibfield  {author} {\bibinfo {author} {\bibfnamefont {A.}~\bibnamefont
  {Puecher}}, \bibinfo {author} {\bibfnamefont {C.}~\bibnamefont {Kalaghatgi}},
  \bibinfo {author} {\bibfnamefont {S.}~\bibnamefont {Roy}}, \bibinfo {author}
  {\bibfnamefont {Y.}~\bibnamefont {Setyawati}}, \bibinfo {author}
  {\bibfnamefont {I.}~\bibnamefont {Gupta}}, \bibinfo {author} {\bibfnamefont
  {B.~S.}\ \bibnamefont {Sathyaprakash}}, \ and\ \bibinfo {author}
  {\bibfnamefont {C.}~\bibnamefont {Van Den~Broeck}},\ }\href {\doibase
  10.1103/PhysRevD.106.082003} {\bibfield  {journal} {\bibinfo  {journal}
  {Phys. Rev. D}\ }\textbf {\bibinfo {volume} {106}},\ \bibinfo {pages}
  {082003} (\bibinfo {year} {2022})},\ \Eprint
  {http://arxiv.org/abs/2205.09062} {arXiv:2205.09062 [gr-qc]} \BibitemShut
  {NoStop}%
\bibitem [{\citenamefont {Mahapatra}(2024)}]{Mahapatra:2023hqq}%
  \BibitemOpen
  \bibfield  {author} {\bibinfo {author} {\bibfnamefont {P.}~\bibnamefont
  {Mahapatra}},\ }\href {\doibase 10.1103/PhysRevD.109.024050} {\bibfield
  {journal} {\bibinfo  {journal} {Phys. Rev. D}\ }\textbf {\bibinfo {volume}
  {109}},\ \bibinfo {pages} {024050} (\bibinfo {year} {2024})},\ \Eprint
  {http://arxiv.org/abs/2306.04703} {arXiv:2306.04703 [gr-qc]} \BibitemShut
  {NoStop}%
\bibitem [{\citenamefont {Mahapatra}\ and\ \citenamefont
  {Kastha}(2024)}]{Mahapatra:2023ydi}%
  \BibitemOpen
  \bibfield  {author} {\bibinfo {author} {\bibfnamefont {P.}~\bibnamefont
  {Mahapatra}}\ and\ \bibinfo {author} {\bibfnamefont {S.}~\bibnamefont
  {Kastha}},\ }\href {\doibase 10.1103/PhysRevD.109.084069} {\bibfield
  {journal} {\bibinfo  {journal} {Phys. Rev. D}\ }\textbf {\bibinfo {volume}
  {109}},\ \bibinfo {pages} {084069} (\bibinfo {year} {2024})},\ \Eprint
  {http://arxiv.org/abs/2311.04672} {arXiv:2311.04672 [gr-qc]} \BibitemShut
  {NoStop}%
\bibitem [{\citenamefont {Blanchet}\ and\ \citenamefont
  {Sathyaprakash}(1994)}]{BSat94}%
  \BibitemOpen
  \bibfield  {author} {\bibinfo {author} {\bibfnamefont {L.}~\bibnamefont
  {Blanchet}}\ and\ \bibinfo {author} {\bibfnamefont {B.~S.}\ \bibnamefont
  {Sathyaprakash}},\ }\href@noop {} {\bibfield  {journal} {\bibinfo  {journal}
  {Class. Quantum Grav.}\ }\textbf {\bibinfo {volume} {11}},\ \bibinfo {pages}
  {2807} (\bibinfo {year} {1994})}\BibitemShut {NoStop}%
\bibitem [{\citenamefont {Blanchet}\ and\ \citenamefont
  {Sathyaprakash}(1995)}]{BSat95}%
  \BibitemOpen
  \bibfield  {author} {\bibinfo {author} {\bibfnamefont {L.}~\bibnamefont
  {Blanchet}}\ and\ \bibinfo {author} {\bibfnamefont {B.~S.}\ \bibnamefont
  {Sathyaprakash}},\ }\href@noop {} {\bibfield  {journal} {\bibinfo  {journal}
  {Phys. Rev. Lett.}\ }\textbf {\bibinfo {volume} {74}},\ \bibinfo {pages}
  {1067} (\bibinfo {year} {1995})}\BibitemShut {NoStop}%
\bibitem [{\citenamefont {Arun}\ \emph
  {et~al.}(2006{\natexlab{b}})\citenamefont {Arun}, \citenamefont {Iyer},
  \citenamefont {Qusailah},\ and\ \citenamefont {Sathyaprakash}}]{AIQS06b}%
  \BibitemOpen
  \bibfield  {author} {\bibinfo {author} {\bibfnamefont {K.~G.}\ \bibnamefont
  {Arun}}, \bibinfo {author} {\bibfnamefont {B.~R.}\ \bibnamefont {Iyer}},
  \bibinfo {author} {\bibfnamefont {M.~S.~S.}\ \bibnamefont {Qusailah}}, \ and\
  \bibinfo {author} {\bibfnamefont {B.~S.}\ \bibnamefont {Sathyaprakash}},\
  }\href@noop {} {\bibfield  {journal} {\bibinfo  {journal} {Phys.~Rev.~D}\
  }\textbf {\bibinfo {volume} {74}},\ \bibinfo {pages} {024006} (\bibinfo
  {year} {2006}{\natexlab{b}})},\ \Eprint {http://arxiv.org/abs/gr-qc/0604067}
  {gr-qc/0604067} \BibitemShut {NoStop}%
\bibitem [{\citenamefont {Yunes}\ and\ \citenamefont
  {Pretorius}(2009)}]{YunesPretorius09}%
  \BibitemOpen
  \bibfield  {author} {\bibinfo {author} {\bibfnamefont {N.}~\bibnamefont
  {Yunes}}\ and\ \bibinfo {author} {\bibfnamefont {F.}~\bibnamefont
  {Pretorius}},\ }\href {\doibase 10.1103/PhysRevD.80.122003} {\bibfield
  {journal} {\bibinfo  {journal} {Phys. Rev. D}\ }\textbf {\bibinfo {volume}
  {80}},\ \bibinfo {pages} {122003} (\bibinfo {year} {2009})},\ \Eprint
  {http://arxiv.org/abs/0909.3328} {arXiv:0909.3328 [gr-qc]} \BibitemShut
  {NoStop}%
\bibitem [{\citenamefont {Mishra}\ \emph {et~al.}(2010)\citenamefont {Mishra},
  \citenamefont {Arun}, \citenamefont {Iyer},\ and\ \citenamefont
  {Sathyaprakash}}]{MAIS10}%
  \BibitemOpen
  \bibfield  {author} {\bibinfo {author} {\bibfnamefont {C.~K.}\ \bibnamefont
  {Mishra}}, \bibinfo {author} {\bibfnamefont {K.~G.}\ \bibnamefont {Arun}},
  \bibinfo {author} {\bibfnamefont {B.~R.}\ \bibnamefont {Iyer}}, \ and\
  \bibinfo {author} {\bibfnamefont {B.~S.}\ \bibnamefont {Sathyaprakash}},\
  }\href@noop {} {\bibfield  {journal} {\bibinfo  {journal} {Phys. Rev. {\bf
  D}}\ }\textbf {\bibinfo {volume} {82}},\ \bibinfo {pages} {064010} (\bibinfo
  {year} {2010})},\ \Eprint {http://arxiv.org/abs/1005.0304} {arXiv:1005.0304
  [gr-qc]} \BibitemShut {NoStop}%
\bibitem [{\citenamefont {Li}\ \emph {et~al.}(2012)\citenamefont {Li},
  \citenamefont {Del~Pozzo}, \citenamefont {Vitale}, \citenamefont {Van
  Den~Broeck}, \citenamefont {Agathos}, \citenamefont {Veitch}, \citenamefont
  {Grover}, \citenamefont {Sidery}, \citenamefont {Sturani},\ and\
  \citenamefont {Vecchio}}]{Li:2011cg}%
  \BibitemOpen
  \bibfield  {author} {\bibinfo {author} {\bibfnamefont {T.~G.~F.}\
  \bibnamefont {Li}}, \bibinfo {author} {\bibfnamefont {W.}~\bibnamefont
  {Del~Pozzo}}, \bibinfo {author} {\bibfnamefont {S.}~\bibnamefont {Vitale}},
  \bibinfo {author} {\bibfnamefont {C.}~\bibnamefont {Van Den~Broeck}},
  \bibinfo {author} {\bibfnamefont {M.}~\bibnamefont {Agathos}}, \bibinfo
  {author} {\bibfnamefont {J.}~\bibnamefont {Veitch}}, \bibinfo {author}
  {\bibfnamefont {K.}~\bibnamefont {Grover}}, \bibinfo {author} {\bibfnamefont
  {T.}~\bibnamefont {Sidery}}, \bibinfo {author} {\bibfnamefont
  {R.}~\bibnamefont {Sturani}}, \ and\ \bibinfo {author} {\bibfnamefont
  {A.}~\bibnamefont {Vecchio}},\ }\href {\doibase 10.1103/PhysRevD.85.082003}
  {\bibfield  {journal} {\bibinfo  {journal} {Phys. Rev. D}\ }\textbf {\bibinfo
  {volume} {85}},\ \bibinfo {pages} {082003} (\bibinfo {year} {2012})},\
  \Eprint {http://arxiv.org/abs/1110.0530} {arXiv:1110.0530 [gr-qc]}
  \BibitemShut {NoStop}%
\bibitem [{\citenamefont {Agathos}\ \emph {et~al.}(2014)\citenamefont
  {Agathos}, \citenamefont {Del~Pozzo}, \citenamefont {Li}, \citenamefont
  {Broeck}, \citenamefont {Veitch} \emph {et~al.}}]{TIGER}%
  \BibitemOpen
  \bibfield  {author} {\bibinfo {author} {\bibfnamefont {M.}~\bibnamefont
  {Agathos}}, \bibinfo {author} {\bibfnamefont {W.}~\bibnamefont {Del~Pozzo}},
  \bibinfo {author} {\bibfnamefont {T.~G.~F.}\ \bibnamefont {Li}}, \bibinfo
  {author} {\bibfnamefont {C.~V.~D.}\ \bibnamefont {Broeck}}, \bibinfo {author}
  {\bibfnamefont {J.}~\bibnamefont {Veitch}},  \emph {et~al.},\ }\href
  {\doibase 10.1103/PhysRevD.89.082001} {\bibfield  {journal} {\bibinfo
  {journal} {Phys.Rev.}\ }\textbf {\bibinfo {volume} {D89}},\ \bibinfo {pages}
  {082001} (\bibinfo {year} {2014})},\ \Eprint {http://arxiv.org/abs/1311.0420}
  {arXiv:1311.0420 [gr-qc]} \BibitemShut {NoStop}%
\bibitem [{\citenamefont {Mehta}\ \emph {et~al.}(2022)\citenamefont {Mehta},
  \citenamefont {Buonanno}, \citenamefont {Cotesta}, \citenamefont {Ghosh},
  \citenamefont {Sennett},\ and\ \citenamefont {Steinhoff}}]{Mehta:2022pcn}%
  \BibitemOpen
  \bibfield  {author} {\bibinfo {author} {\bibfnamefont {A.~K.}\ \bibnamefont
  {Mehta}}, \bibinfo {author} {\bibfnamefont {A.}~\bibnamefont {Buonanno}},
  \bibinfo {author} {\bibfnamefont {R.}~\bibnamefont {Cotesta}}, \bibinfo
  {author} {\bibfnamefont {A.}~\bibnamefont {Ghosh}}, \bibinfo {author}
  {\bibfnamefont {N.}~\bibnamefont {Sennett}}, \ and\ \bibinfo {author}
  {\bibfnamefont {J.}~\bibnamefont {Steinhoff}},\ }\href@noop {} {\  (\bibinfo
  {year} {2022})},\ \Eprint {http://arxiv.org/abs/2203.13937} {arXiv:2203.13937
  [gr-qc]} \BibitemShut {NoStop}%
\bibitem [{\citenamefont {Abbott}\ \emph {et~al.}(2016)\citenamefont {Abbott}
  \emph {et~al.}}]{TGR-GW150914}%
  \BibitemOpen
  \bibfield  {author} {\bibinfo {author} {\bibfnamefont {B.~P.}\ \bibnamefont
  {Abbott}} \emph {et~al.} (\bibinfo {collaboration} {LIGO Scientific,
  Virgo}),\ }\href {\doibase 10.1103/PhysRevLett.116.221101} {\bibfield
  {journal} {\bibinfo  {journal} {Phys. Rev. Lett.}\ }\textbf {\bibinfo
  {volume} {116}},\ \bibinfo {pages} {221101} (\bibinfo {year} {2016})},\
  \bibinfo {note} {[Erratum: Phys.Rev.Lett. 121, 129902 (2018)]},\ \Eprint
  {http://arxiv.org/abs/1602.03841} {arXiv:1602.03841 [gr-qc]} \BibitemShut
  {NoStop}%
\bibitem [{\citenamefont {Abbott}\ \emph {et~al.}(2019)\citenamefont {Abbott}
  \emph {et~al.}}]{TGR-GWTC-1}%
  \BibitemOpen
  \bibfield  {author} {\bibinfo {author} {\bibfnamefont {B.~P.}\ \bibnamefont
  {Abbott}} \emph {et~al.} (\bibinfo {collaboration} {LIGO Scientific,
  Virgo}),\ }\href {\doibase 10.1103/PhysRevD.100.104036} {\bibfield  {journal}
  {\bibinfo  {journal} {Phys. Rev. D}\ }\textbf {\bibinfo {volume} {100}},\
  \bibinfo {pages} {104036} (\bibinfo {year} {2019})},\ \Eprint
  {http://arxiv.org/abs/1903.04467} {arXiv:1903.04467 [gr-qc]} \BibitemShut
  {NoStop}%
\bibitem [{\citenamefont {Abbott}\ \emph
  {et~al.}(2021{\natexlab{a}})\citenamefont {Abbott} \emph
  {et~al.}}]{TGR-GWTC-2}%
  \BibitemOpen
  \bibfield  {author} {\bibinfo {author} {\bibfnamefont {R.}~\bibnamefont
  {Abbott}} \emph {et~al.} (\bibinfo {collaboration} {LIGO Scientific,
  Virgo}),\ }\href {\doibase 10.1103/PhysRevD.103.122002} {\bibfield  {journal}
  {\bibinfo  {journal} {Phys. Rev. D}\ }\textbf {\bibinfo {volume} {103}},\
  \bibinfo {pages} {122002} (\bibinfo {year} {2021}{\natexlab{a}})},\ \Eprint
  {http://arxiv.org/abs/2010.14529} {arXiv:2010.14529 [gr-qc]} \BibitemShut
  {NoStop}%
\bibitem [{\citenamefont {Abbott}\ \emph
  {et~al.}(2021{\natexlab{b}})\citenamefont {Abbott} \emph
  {et~al.}}]{TGR-GWTC-3}%
  \BibitemOpen
  \bibfield  {author} {\bibinfo {author} {\bibfnamefont {R.}~\bibnamefont
  {Abbott}} \emph {et~al.} (\bibinfo {collaboration} {LIGO Scientific, VIRGO,
  KAGRA}),\ }\href@noop {} {\  (\bibinfo {year} {2021}{\natexlab{b}})},\
  \Eprint {http://arxiv.org/abs/2112.06861} {arXiv:2112.06861 [gr-qc]}
  \BibitemShut {NoStop}%
\bibitem [{\citenamefont {Pai}\ and\ \citenamefont {Arun}(2013)}]{AP12}%
  \BibitemOpen
  \bibfield  {author} {\bibinfo {author} {\bibfnamefont {A.}~\bibnamefont
  {Pai}}\ and\ \bibinfo {author} {\bibfnamefont {K.}~\bibnamefont {Arun}},\
  }\href {\doibase 10.1088/0264-9381/30/2/025011} {\bibfield  {journal}
  {\bibinfo  {journal} {Class.Quant.Grav.}\ }\textbf {\bibinfo {volume} {30}},\
  \bibinfo {pages} {025011} (\bibinfo {year} {2013})},\ \Eprint
  {http://arxiv.org/abs/1207.1943} {arXiv:1207.1943 [gr-qc]} \BibitemShut
  {NoStop}%
\bibitem [{\citenamefont {Datta}\ \emph {et~al.}(2022)\citenamefont {Datta},
  \citenamefont {Saleem}, \citenamefont {Arun},\ and\ \citenamefont
  {Sathyaprakash}}]{Datta:2022izc}%
  \BibitemOpen
  \bibfield  {author} {\bibinfo {author} {\bibfnamefont {S.}~\bibnamefont
  {Datta}}, \bibinfo {author} {\bibfnamefont {M.}~\bibnamefont {Saleem}},
  \bibinfo {author} {\bibfnamefont {K.~G.}\ \bibnamefont {Arun}}, \ and\
  \bibinfo {author} {\bibfnamefont {B.~S.}\ \bibnamefont {Sathyaprakash}},\
  }\href@noop {} {\  (\bibinfo {year} {2022})},\ \Eprint
  {http://arxiv.org/abs/2208.07757} {arXiv:2208.07757 [gr-qc]} \BibitemShut
  {NoStop}%
\bibitem [{\citenamefont {Datta}(2023)}]{Datta:2023muk}%
  \BibitemOpen
  \bibfield  {author} {\bibinfo {author} {\bibfnamefont {S.}~\bibnamefont
  {Datta}},\ }\href@noop {} {\  (\bibinfo {year} {2023})},\ \Eprint
  {http://arxiv.org/abs/2303.04399} {arXiv:2303.04399 [gr-qc]} \BibitemShut
  {NoStop}%
\bibitem [{\citenamefont {Yunes}\ \emph {et~al.}(2016)\citenamefont {Yunes},
  \citenamefont {Yagi},\ and\ \citenamefont {Pretorius}}]{YYP2016}%
  \BibitemOpen
  \bibfield  {author} {\bibinfo {author} {\bibfnamefont {N.}~\bibnamefont
  {Yunes}}, \bibinfo {author} {\bibfnamefont {K.}~\bibnamefont {Yagi}}, \ and\
  \bibinfo {author} {\bibfnamefont {F.}~\bibnamefont {Pretorius}},\ }\href@noop
  {} {\  (\bibinfo {year} {2016})},\ \Eprint {http://arxiv.org/abs/1603.08955}
  {arXiv:1603.08955 [gr-qc]} \BibitemShut {NoStop}%
\bibitem [{\citenamefont {Van Den~Broeck}\ and\ \citenamefont
  {Sengupta}(2007)}]{VanDenBroeck:2006ar}%
  \BibitemOpen
  \bibfield  {author} {\bibinfo {author} {\bibfnamefont {C.}~\bibnamefont {Van
  Den~Broeck}}\ and\ \bibinfo {author} {\bibfnamefont {A.~S.}\ \bibnamefont
  {Sengupta}},\ }\href {\doibase 10.1088/0264-9381/24/5/005} {\bibfield
  {journal} {\bibinfo  {journal} {Class. Quant. Grav.}\ }\textbf {\bibinfo
  {volume} {24}},\ \bibinfo {pages} {1089} (\bibinfo {year} {2007})},\ \Eprint
  {http://arxiv.org/abs/gr-qc/0610126} {arXiv:gr-qc/0610126} \BibitemShut
  {NoStop}%
\bibitem [{\citenamefont {{Arun}}\ \emph {et~al.}(2007)\citenamefont {{Arun}},
  \citenamefont {{Iyer}}, \citenamefont {{Sathyaprakash}},\ and\ \citenamefont
  {{Sinha}}}]{AISS07}%
  \BibitemOpen
  \bibfield  {author} {\bibinfo {author} {\bibfnamefont {K.~G.}\ \bibnamefont
  {{Arun}}}, \bibinfo {author} {\bibfnamefont {B.~R.}\ \bibnamefont {{Iyer}}},
  \bibinfo {author} {\bibfnamefont {B.~S.}\ \bibnamefont {{Sathyaprakash}}}, \
  and\ \bibinfo {author} {\bibfnamefont {S.}~\bibnamefont {{Sinha}}},\
  }\href@noop {} {\bibfield  {journal} {\bibinfo  {journal} {Phys.~Rev.~D}\
  }\textbf {\bibinfo {volume} {75}},\ \bibinfo {pages} {124002} (\bibinfo
  {year} {2007})},\ \Eprint {http://arxiv.org/abs/0704.1086} {0704.1086}
  \BibitemShut {NoStop}%
\bibitem [{\citenamefont {Arun}\ \emph {et~al.}(2009)\citenamefont {Arun},
  \citenamefont {Buonanno}, \citenamefont {Faye},\ and\ \citenamefont
  {Ochsner}}]{ABFO08}%
  \BibitemOpen
  \bibfield  {author} {\bibinfo {author} {\bibfnamefont {K.~G.}\ \bibnamefont
  {Arun}}, \bibinfo {author} {\bibfnamefont {A.}~\bibnamefont {Buonanno}},
  \bibinfo {author} {\bibfnamefont {G.}~\bibnamefont {Faye}}, \ and\ \bibinfo
  {author} {\bibfnamefont {E.}~\bibnamefont {Ochsner}},\ }\href@noop {}
  {\bibfield  {journal} {\bibinfo  {journal} {Phys. Rev. D}\ }\textbf {\bibinfo
  {volume} {79}},\ \bibinfo {pages} {104023} (\bibinfo {year} {2009})},\
  \Eprint {http://arxiv.org/abs/0810.5336} {arXiv:0810.5336 [gr-qc]}
  \BibitemShut {NoStop}%
\bibitem [{\citenamefont {Borhanian}(2021)}]{gwbench}%
  \BibitemOpen
  \bibfield  {author} {\bibinfo {author} {\bibfnamefont {S.}~\bibnamefont
  {Borhanian}},\ }\href {\doibase 10.1088/1361-6382/ac1618} {\bibfield
  {journal} {\bibinfo  {journal} {Class. Quant. Grav.}\ }\textbf {\bibinfo
  {volume} {38}},\ \bibinfo {pages} {175014} (\bibinfo {year} {2021})},\
  \Eprint {http://arxiv.org/abs/2010.15202} {arXiv:2010.15202 [gr-qc]}
  \BibitemShut {NoStop}%
\bibitem [{\citenamefont {Fritschel}\ \emph {et~al.}(2022)\citenamefont
  {Fritschel}, \citenamefont {Kuns}, \citenamefont {Driggers}, \citenamefont
  {Effler}, \citenamefont {Lantz}, \citenamefont {Ottaway}, \citenamefont
  {Ballmer}, \citenamefont {Dooley}, \citenamefont {Adhikari}, \citenamefont
  {Evans}, \citenamefont {Farr}, \citenamefont {Gonzalez}, \citenamefont
  {Schmidt},\ and\ \citenamefont {Raja}}]{T2200287}%
  \BibitemOpen
  \bibfield  {author} {\bibinfo {author} {\bibfnamefont {P.}~\bibnamefont
  {Fritschel}}, \bibinfo {author} {\bibfnamefont {K.}~\bibnamefont {Kuns}},
  \bibinfo {author} {\bibfnamefont {J.}~\bibnamefont {Driggers}}, \bibinfo
  {author} {\bibfnamefont {A.}~\bibnamefont {Effler}}, \bibinfo {author}
  {\bibfnamefont {B.}~\bibnamefont {Lantz}}, \bibinfo {author} {\bibfnamefont
  {D.}~\bibnamefont {Ottaway}}, \bibinfo {author} {\bibfnamefont
  {S.}~\bibnamefont {Ballmer}}, \bibinfo {author} {\bibfnamefont
  {K.}~\bibnamefont {Dooley}}, \bibinfo {author} {\bibfnamefont
  {R.}~\bibnamefont {Adhikari}}, \bibinfo {author} {\bibfnamefont
  {M.}~\bibnamefont {Evans}}, \bibinfo {author} {\bibfnamefont
  {B.}~\bibnamefont {Farr}}, \bibinfo {author} {\bibfnamefont {G.}~\bibnamefont
  {Gonzalez}}, \bibinfo {author} {\bibfnamefont {P.}~\bibnamefont {Schmidt}}, \
  and\ \bibinfo {author} {\bibfnamefont {S.}~\bibnamefont {Raja}},\ }\href
  {https://dcc.ligo.org/LIGO-T2200287/public} {\emph {\bibinfo {title} {Report
  from the LSC Post-O5 Study Group}}},\ \bibinfo {type} {Tech. Rep.}\ \bibinfo
  {number} {T2200287}\ (\bibinfo  {institution} {LIGO},\ \bibinfo {year}
  {2022})\BibitemShut {NoStop}%
\bibitem [{\citenamefont {Iyer}\ \emph {et~al.}(2011)\citenamefont {Iyer},
  \citenamefont {Souradeep}, \citenamefont {Unnikrishnan}, \citenamefont
  {Dhurandhar}, \citenamefont {Raja},\ and\ \citenamefont
  {Sengupta}}]{LIGO-India}%
  \BibitemOpen
  \bibfield  {author} {\bibinfo {author} {\bibfnamefont {B.}~\bibnamefont
  {Iyer}}, \bibinfo {author} {\bibfnamefont {T.}~\bibnamefont {Souradeep}},
  \bibinfo {author} {\bibfnamefont {C.}~\bibnamefont {Unnikrishnan}}, \bibinfo
  {author} {\bibfnamefont {S.}~\bibnamefont {Dhurandhar}}, \bibinfo {author}
  {\bibfnamefont {S.}~\bibnamefont {Raja}}, \ and\ \bibinfo {author}
  {\bibfnamefont {A.}~\bibnamefont {Sengupta}},\ }\href
  {https://dcc.ligo.org/LIGO-M1100296/public} {\emph {\bibinfo {title}
  {LIGO-India, Proposal of the Consortium for Indian Initiative in
  Gravitational-wave Observations (IndIGO)}}},\ \bibinfo {type} {Tech. Rep.}\
  \bibinfo {number} {M1100296-v2}\ (\bibinfo  {institution} {LIGO-India},\
  \bibinfo {year} {2011})\BibitemShut {NoStop}%
\bibitem [{\citenamefont {Evans}\ \emph {et~al.}(2021)\citenamefont {Evans}
  \emph {et~al.}}]{Evans:2021gyd}%
  \BibitemOpen
  \bibfield  {author} {\bibinfo {author} {\bibfnamefont {M.}~\bibnamefont
  {Evans}} \emph {et~al.},\ }\href@noop {} {\  (\bibinfo {year} {2021})},\
  \Eprint {http://arxiv.org/abs/2109.09882} {arXiv:2109.09882 [astro-ph.IM]}
  \BibitemShut {NoStop}%
\bibitem [{\citenamefont {Hild}\ \emph {et~al.}(2011)\citenamefont {Hild} \emph
  {et~al.}}]{Hild:2010id}%
  \BibitemOpen
  \bibfield  {author} {\bibinfo {author} {\bibfnamefont {S.}~\bibnamefont
  {Hild}} \emph {et~al.},\ }\href {\doibase 10.1088/0264-9381/28/9/094013}
  {\bibfield  {journal} {\bibinfo  {journal} {Class. Quant. Grav.}\ }\textbf
  {\bibinfo {volume} {28}},\ \bibinfo {pages} {094013} (\bibinfo {year}
  {2011})},\ \Eprint {http://arxiv.org/abs/1012.0908} {arXiv:1012.0908 [gr-qc]}
  \BibitemShut {NoStop}%
\bibitem [{\citenamefont {Punturo}\ \emph {et~al.}(2010)\citenamefont {Punturo}
  \emph {et~al.}}]{Punturo:2010zz}%
  \BibitemOpen
  \bibfield  {author} {\bibinfo {author} {\bibfnamefont {M.}~\bibnamefont
  {Punturo}} \emph {et~al.},\ }\href {\doibase 10.1088/0264-9381/27/19/194002}
  {\bibfield  {journal} {\bibinfo  {journal} {Class. Quant. Grav.}\ }\textbf
  {\bibinfo {volume} {27}},\ \bibinfo {pages} {194002} (\bibinfo {year}
  {2010})}\BibitemShut {NoStop}%
\bibitem [{\citenamefont {Branchesi}\ \emph {et~al.}(2023)\citenamefont
  {Branchesi} \emph {et~al.}}]{Branchesi:2023mws}%
  \BibitemOpen
  \bibfield  {author} {\bibinfo {author} {\bibfnamefont {M.}~\bibnamefont
  {Branchesi}} \emph {et~al.},\ }\href {\doibase 10.1088/1475-7516/2023/07/068}
  {\bibfield  {journal} {\bibinfo  {journal} {JCAP}\ }\textbf {\bibinfo
  {volume} {07}},\ \bibinfo {pages} {068} (\bibinfo {year} {2023})},\ \Eprint
  {http://arxiv.org/abs/2303.15923} {arXiv:2303.15923 [gr-qc]} \BibitemShut
  {NoStop}%
\bibitem [{\citenamefont {{Gupta}}\ \emph {et~al.}(2023)\citenamefont
  {{Gupta}}, \citenamefont {{Afle}}, \citenamefont {{Arun}} \emph
  {et~al.}}]{Gupta2023mpsac}%
  \BibitemOpen
  \bibfield  {author} {\bibinfo {author} {\bibfnamefont {I.}~\bibnamefont
  {{Gupta}}}, \bibinfo {author} {\bibfnamefont {C.}~\bibnamefont {{Afle}}},
  \bibinfo {author} {\bibfnamefont {K.~G.}\ \bibnamefont {{Arun}}},  \emph
  {et~al.},\ }\href {\doibase 10.48550/arXiv.2307.10421} {\bibfield  {journal}
  {\bibinfo  {journal} {arXiv e-prints}\ ,\ \bibinfo {eid} {arXiv:2307.10421}}
  (\bibinfo {year} {2023})},\ \Eprint {http://arxiv.org/abs/2307.10421}
  {arXiv:2307.10421 [gr-qc]} \BibitemShut {NoStop}%
\bibitem [{\citenamefont {Rao}(1945)}]{Rao45}%
  \BibitemOpen
  \bibfield  {author} {\bibinfo {author} {\bibfnamefont {C.}~\bibnamefont
  {Rao}},\ }\href@noop {} {\bibfield  {journal} {\bibinfo  {journal} {Bullet.
  Calcutta Math. Soc}\ }\textbf {\bibinfo {volume} {37}},\ \bibinfo {pages}
  {81} (\bibinfo {year} {1945})}\BibitemShut {NoStop}%
\bibitem [{\citenamefont {Cramer}(1946)}]{Cramer46}%
  \BibitemOpen
  \bibfield  {author} {\bibinfo {author} {\bibfnamefont {H.}~\bibnamefont
  {Cramer}},\ }\href@noop {} {\emph {\bibinfo {title} {Mathematical methods in
  statistics}}}\ (\bibinfo  {publisher} {Pergamon Press},\ \bibinfo {address}
  {Princeton University Press, NJ, U.S.A.},\ \bibinfo {year}
  {1946})\BibitemShut {NoStop}%
\bibitem [{\citenamefont {Cutler}\ and\ \citenamefont {Flanagan}(1994)}]{CF94}%
  \BibitemOpen
  \bibfield  {author} {\bibinfo {author} {\bibfnamefont {C.}~\bibnamefont
  {Cutler}}\ and\ \bibinfo {author} {\bibfnamefont {E.}~\bibnamefont
  {Flanagan}},\ }\href@noop {} {\bibfield  {journal} {\bibinfo  {journal}
  {Phys. Rev. D}\ }\textbf {\bibinfo {volume} {49}},\ \bibinfo {pages} {2658}
  (\bibinfo {year} {1994})}\BibitemShut {NoStop}%
\bibitem [{\citenamefont {Poisson}\ and\ \citenamefont {Will}(1995)}]{PW95}%
  \BibitemOpen
  \bibfield  {author} {\bibinfo {author} {\bibfnamefont {E.}~\bibnamefont
  {Poisson}}\ and\ \bibinfo {author} {\bibfnamefont {C.}~\bibnamefont {Will}},\
  }\href@noop {} {\bibfield  {journal} {\bibinfo  {journal} {Phys. Rev. D}\
  }\textbf {\bibinfo {volume} {52}},\ \bibinfo {pages} {848} (\bibinfo {year}
  {1995})}\BibitemShut {NoStop}%
\bibitem [{\citenamefont {{Bardeen}}\ \emph {et~al.}(1972)\citenamefont
  {{Bardeen}}, \citenamefont {{Press}},\ and\ \citenamefont
  {{Teukolsky}}}]{Bardeen72}%
  \BibitemOpen
  \bibfield  {author} {\bibinfo {author} {\bibfnamefont {J.~M.}\ \bibnamefont
  {{Bardeen}}}, \bibinfo {author} {\bibfnamefont {W.~H.}\ \bibnamefont
  {{Press}}}, \ and\ \bibinfo {author} {\bibfnamefont {S.~A.}\ \bibnamefont
  {{Teukolsky}}},\ }\href {\doibase 10.1086/151796} {\bibfield  {journal}
  {\bibinfo  {journal} {\apj}\ }\textbf {\bibinfo {volume} {178}},\ \bibinfo
  {pages} {347} (\bibinfo {year} {1972})}\BibitemShut {NoStop}%
\bibitem [{\citenamefont {Husa}\ \emph {et~al.}(2016)\citenamefont {Husa},
  \citenamefont {Khan}, \citenamefont {Hannam}, \citenamefont {P\"urrer},
  \citenamefont {Ohme}, \citenamefont {Jim\'enez~Forteza},\ and\ \citenamefont
  {Boh\'e}}]{Husa:2015iqa}%
  \BibitemOpen
  \bibfield  {author} {\bibinfo {author} {\bibfnamefont {S.}~\bibnamefont
  {Husa}}, \bibinfo {author} {\bibfnamefont {S.}~\bibnamefont {Khan}}, \bibinfo
  {author} {\bibfnamefont {M.}~\bibnamefont {Hannam}}, \bibinfo {author}
  {\bibfnamefont {M.}~\bibnamefont {P\"urrer}}, \bibinfo {author}
  {\bibfnamefont {F.}~\bibnamefont {Ohme}}, \bibinfo {author} {\bibfnamefont
  {X.}~\bibnamefont {Jim\'enez~Forteza}}, \ and\ \bibinfo {author}
  {\bibfnamefont {A.}~\bibnamefont {Boh\'e}},\ }\href {\doibase
  10.1103/PhysRevD.93.044006} {\bibfield  {journal} {\bibinfo  {journal} {Phys.
  Rev. D}\ }\textbf {\bibinfo {volume} {93}},\ \bibinfo {pages} {044006}
  (\bibinfo {year} {2016})},\ \Eprint {http://arxiv.org/abs/1508.07250}
  {arXiv:1508.07250 [gr-qc]} \BibitemShut {NoStop}%
\bibitem [{\citenamefont {Hofmann}\ \emph {et~al.}(2016)\citenamefont
  {Hofmann}, \citenamefont {Barausse},\ and\ \citenamefont
  {Rezzolla}}]{Hofmann:2016yih}%
  \BibitemOpen
  \bibfield  {author} {\bibinfo {author} {\bibfnamefont {F.}~\bibnamefont
  {Hofmann}}, \bibinfo {author} {\bibfnamefont {E.}~\bibnamefont {Barausse}}, \
  and\ \bibinfo {author} {\bibfnamefont {L.}~\bibnamefont {Rezzolla}},\ }\href
  {\doibase 10.3847/2041-8205/825/2/L19} {\bibfield  {journal} {\bibinfo
  {journal} {Astrophys. J. Lett.}\ }\textbf {\bibinfo {volume} {825}},\
  \bibinfo {pages} {L19} (\bibinfo {year} {2016})},\ \Eprint
  {http://arxiv.org/abs/1605.01938} {arXiv:1605.01938 [gr-qc]} \BibitemShut
  {NoStop}%
\bibitem [{\citenamefont {Favata}\ \emph {et~al.}(2022)\citenamefont {Favata},
  \citenamefont {Kim}, \citenamefont {Arun}, \citenamefont {Kim},\ and\
  \citenamefont {Lee}}]{Favata:2021vhw}%
  \BibitemOpen
  \bibfield  {author} {\bibinfo {author} {\bibfnamefont {M.}~\bibnamefont
  {Favata}}, \bibinfo {author} {\bibfnamefont {C.}~\bibnamefont {Kim}},
  \bibinfo {author} {\bibfnamefont {K.~G.}\ \bibnamefont {Arun}}, \bibinfo
  {author} {\bibfnamefont {J.}~\bibnamefont {Kim}}, \ and\ \bibinfo {author}
  {\bibfnamefont {H.~W.}\ \bibnamefont {Lee}},\ }\href {\doibase
  10.1103/PhysRevD.105.023003} {\bibfield  {journal} {\bibinfo  {journal}
  {Phys. Rev. D}\ }\textbf {\bibinfo {volume} {105}},\ \bibinfo {pages}
  {023003} (\bibinfo {year} {2022})},\ \Eprint
  {http://arxiv.org/abs/2108.05861} {arXiv:2108.05861 [gr-qc]} \BibitemShut
  {NoStop}%
\bibitem [{\citenamefont {{Berti}}\ \emph {et~al.}(2005)\citenamefont
  {{Berti}}, \citenamefont {{Buonanno}},\ and\ \citenamefont
  {{Will}}}]{BBW05a}%
  \BibitemOpen
  \bibfield  {author} {\bibinfo {author} {\bibfnamefont {E.}~\bibnamefont
  {{Berti}}}, \bibinfo {author} {\bibfnamefont {A.}~\bibnamefont {{Buonanno}}},
  \ and\ \bibinfo {author} {\bibfnamefont {C.~M.}\ \bibnamefont {{Will}}},\
  }\href {\doibase 10.1103/PhysRevD.71.084025} {\bibfield  {journal} {\bibinfo
  {journal} {Phys.~Rev.~D}\ }\textbf {\bibinfo {volume} {71}},\ \bibinfo
  {pages} {084025} (\bibinfo {year} {2005})},\ \Eprint
  {http://arxiv.org/abs/gr-qc/0411129} {gr-qc/0411129} \BibitemShut {NoStop}%
\bibitem [{\citenamefont {Mangiagli}\ \emph {et~al.}(2020)\citenamefont
  {Mangiagli}, \citenamefont {Klein}, \citenamefont {Bonetti}, \citenamefont
  {Katz}, \citenamefont {Sesana}, \citenamefont {Volonteri}, \citenamefont
  {Colpi}, \citenamefont {Marsat},\ and\ \citenamefont
  {Babak}}]{Mangiagli:2020rwz}%
  \BibitemOpen
  \bibfield  {author} {\bibinfo {author} {\bibfnamefont {A.}~\bibnamefont
  {Mangiagli}}, \bibinfo {author} {\bibfnamefont {A.}~\bibnamefont {Klein}},
  \bibinfo {author} {\bibfnamefont {M.}~\bibnamefont {Bonetti}}, \bibinfo
  {author} {\bibfnamefont {M.~L.}\ \bibnamefont {Katz}}, \bibinfo {author}
  {\bibfnamefont {A.}~\bibnamefont {Sesana}}, \bibinfo {author} {\bibfnamefont
  {M.}~\bibnamefont {Volonteri}}, \bibinfo {author} {\bibfnamefont
  {M.}~\bibnamefont {Colpi}}, \bibinfo {author} {\bibfnamefont
  {S.}~\bibnamefont {Marsat}}, \ and\ \bibinfo {author} {\bibfnamefont
  {S.}~\bibnamefont {Babak}},\ }\href {\doibase 10.1103/PhysRevD.102.084056}
  {\bibfield  {journal} {\bibinfo  {journal} {Phys. Rev. D}\ }\textbf {\bibinfo
  {volume} {102}},\ \bibinfo {pages} {084056} (\bibinfo {year} {2020})},\
  \Eprint {http://arxiv.org/abs/2006.12513} {arXiv:2006.12513 [astro-ph.HE]}
  \BibitemShut {NoStop}%
\bibitem [{\citenamefont {Abbott}\ \emph
  {et~al.}(2020{\natexlab{a}})\citenamefont {Abbott} \emph
  {et~al.}}]{GW190412}%
  \BibitemOpen
  \bibfield  {author} {\bibinfo {author} {\bibfnamefont {R.}~\bibnamefont
  {Abbott}} \emph {et~al.} (\bibinfo {collaboration} {LIGO Scientific,
  Virgo}),\ }\href {\doibase 10.1103/PhysRevD.102.043015} {\bibfield  {journal}
  {\bibinfo  {journal} {Phys. Rev. D}\ }\textbf {\bibinfo {volume} {102}},\
  \bibinfo {pages} {043015} (\bibinfo {year} {2020}{\natexlab{a}})},\ \Eprint
  {http://arxiv.org/abs/2004.08342} {arXiv:2004.08342 [astro-ph.HE]}
  \BibitemShut {NoStop}%
\bibitem [{\citenamefont {Abbott}\ \emph
  {et~al.}(2020{\natexlab{b}})\citenamefont {Abbott} \emph
  {et~al.}}]{GW190814}%
  \BibitemOpen
  \bibfield  {author} {\bibinfo {author} {\bibfnamefont {R.}~\bibnamefont
  {Abbott}} \emph {et~al.} (\bibinfo {collaboration} {LIGO Scientific,
  Virgo}),\ }\href {\doibase 10.3847/2041-8213/ab960f} {\bibfield  {journal}
  {\bibinfo  {journal} {Astrophys. J. Lett.}\ }\textbf {\bibinfo {volume}
  {896}},\ \bibinfo {pages} {L44} (\bibinfo {year} {2020}{\natexlab{b}})},\
  \Eprint {http://arxiv.org/abs/2006.12611} {arXiv:2006.12611 [astro-ph.HE]}
  \BibitemShut {NoStop}%
\bibitem [{\citenamefont {Abbott}\ \emph
  {et~al.}(2023{\natexlab{a}})\citenamefont {Abbott} \emph
  {et~al.}}]{KAGRA:2023pio}%
  \BibitemOpen
  \bibfield  {author} {\bibinfo {author} {\bibfnamefont {R.}~\bibnamefont
  {Abbott}} \emph {et~al.} (\bibinfo {collaboration} {KAGRA, VIRGO, LIGO
  Scientific}),\ }\href {\doibase 10.3847/1538-4365/acdc9f} {\bibfield
  {journal} {\bibinfo  {journal} {Astrophys. J. Suppl.}\ }\textbf {\bibinfo
  {volume} {267}},\ \bibinfo {pages} {29} (\bibinfo {year}
  {2023}{\natexlab{a}})},\ \Eprint {http://arxiv.org/abs/2302.03676}
  {arXiv:2302.03676 [gr-qc]} \BibitemShut {NoStop}%
\bibitem [{\citenamefont {Krishnendu}\ \emph {et~al.}(2017)\citenamefont
  {Krishnendu}, \citenamefont {Arun},\ and\ \citenamefont
  {Mishra}}]{Krishnendu:2017shb}%
  \BibitemOpen
  \bibfield  {author} {\bibinfo {author} {\bibfnamefont {N.~V.}\ \bibnamefont
  {Krishnendu}}, \bibinfo {author} {\bibfnamefont {K.~G.}\ \bibnamefont
  {Arun}}, \ and\ \bibinfo {author} {\bibfnamefont {C.~K.}\ \bibnamefont
  {Mishra}},\ }\href {\doibase 10.1103/PhysRevLett.119.091101} {\bibfield
  {journal} {\bibinfo  {journal} {Phys. Rev. Lett.}\ }\textbf {\bibinfo
  {volume} {119}},\ \bibinfo {pages} {091101} (\bibinfo {year} {2017})},\
  \Eprint {http://arxiv.org/abs/1701.06318} {arXiv:1701.06318 [gr-qc]}
  \BibitemShut {NoStop}%
\bibitem [{\citenamefont {Krishnendu}\ \emph
  {et~al.}(2019{\natexlab{a}})\citenamefont {Krishnendu}, \citenamefont
  {Mishra},\ and\ \citenamefont {Arun}}]{Krishnendu:2018nqa}%
  \BibitemOpen
  \bibfield  {author} {\bibinfo {author} {\bibfnamefont {N.~V.}\ \bibnamefont
  {Krishnendu}}, \bibinfo {author} {\bibfnamefont {C.~K.}\ \bibnamefont
  {Mishra}}, \ and\ \bibinfo {author} {\bibfnamefont {K.~G.}\ \bibnamefont
  {Arun}},\ }\href {\doibase 10.1103/PhysRevD.99.064008} {\bibfield  {journal}
  {\bibinfo  {journal} {Phys. Rev. D}\ }\textbf {\bibinfo {volume} {99}},\
  \bibinfo {pages} {064008} (\bibinfo {year} {2019}{\natexlab{a}})},\ \Eprint
  {http://arxiv.org/abs/1811.00317} {arXiv:1811.00317 [gr-qc]} \BibitemShut
  {NoStop}%
\bibitem [{\citenamefont {Krishnendu}\ \emph
  {et~al.}(2019{\natexlab{b}})\citenamefont {Krishnendu}, \citenamefont
  {Saleem}, \citenamefont {Samajdar}, \citenamefont {Arun}, \citenamefont
  {Del~Pozzo},\ and\ \citenamefont {Mishra}}]{Krishnendu:2019tjp}%
  \BibitemOpen
  \bibfield  {author} {\bibinfo {author} {\bibfnamefont {N.~V.}\ \bibnamefont
  {Krishnendu}}, \bibinfo {author} {\bibfnamefont {M.}~\bibnamefont {Saleem}},
  \bibinfo {author} {\bibfnamefont {A.}~\bibnamefont {Samajdar}}, \bibinfo
  {author} {\bibfnamefont {K.~G.}\ \bibnamefont {Arun}}, \bibinfo {author}
  {\bibfnamefont {W.}~\bibnamefont {Del~Pozzo}}, \ and\ \bibinfo {author}
  {\bibfnamefont {C.~K.}\ \bibnamefont {Mishra}},\ }\href {\doibase
  10.1103/PhysRevD.100.104019} {\bibfield  {journal} {\bibinfo  {journal}
  {Phys. Rev. D}\ }\textbf {\bibinfo {volume} {100}},\ \bibinfo {pages}
  {104019} (\bibinfo {year} {2019}{\natexlab{b}})},\ \Eprint
  {http://arxiv.org/abs/1908.02247} {arXiv:1908.02247 [gr-qc]} \BibitemShut
  {NoStop}%
\bibitem [{\citenamefont {Saini}\ and\ \citenamefont
  {Krishnendu}(2023)}]{Saini:2023gaw}%
  \BibitemOpen
  \bibfield  {author} {\bibinfo {author} {\bibfnamefont {P.}~\bibnamefont
  {Saini}}\ and\ \bibinfo {author} {\bibfnamefont {N.~V.}\ \bibnamefont
  {Krishnendu}},\ }\href@noop {} {\  (\bibinfo {year} {2023})},\ \Eprint
  {http://arxiv.org/abs/2308.01309} {arXiv:2308.01309 [gr-qc]} \BibitemShut
  {NoStop}%
\bibitem [{\citenamefont {Will}(1998)}]{Will98}%
  \BibitemOpen
  \bibfield  {author} {\bibinfo {author} {\bibfnamefont {C.~M.}\ \bibnamefont
  {Will}},\ }\href@noop {} {\bibfield  {journal} {\bibinfo  {journal} {Phys.~
  Rev.~D}\ }\textbf {\bibinfo {volume} {57}},\ \bibinfo {pages} {2061}
  (\bibinfo {year} {1998})},\ \Eprint {http://arxiv.org/abs/gr-qc/9709011}
  {gr-qc/9709011} \BibitemShut {NoStop}%
\bibitem [{\citenamefont {Mirshekari}\ \emph {et~al.}(2012)\citenamefont
  {Mirshekari}, \citenamefont {Yunes},\ and\ \citenamefont {Will}}]{MYW11}%
  \BibitemOpen
  \bibfield  {author} {\bibinfo {author} {\bibfnamefont {S.}~\bibnamefont
  {Mirshekari}}, \bibinfo {author} {\bibfnamefont {N.}~\bibnamefont {Yunes}}, \
  and\ \bibinfo {author} {\bibfnamefont {C.~M.}\ \bibnamefont {Will}},\ }\href
  {\doibase 10.1103/PhysRevD.85.024041} {\bibfield  {journal} {\bibinfo
  {journal} {Phys. Rev. D}\ }\textbf {\bibinfo {volume} {85}},\ \bibinfo
  {pages} {024041} (\bibinfo {year} {2012})},\ \Eprint
  {http://arxiv.org/abs/1110.2720} {arXiv:1110.2720 [gr-qc]} \BibitemShut
  {NoStop}%
\bibitem [{\citenamefont {Kosteleck\'y}\ and\ \citenamefont
  {Mewes}(2016)}]{Kostelecky:2016kfm}%
  \BibitemOpen
  \bibfield  {author} {\bibinfo {author} {\bibfnamefont {V.~A.}\ \bibnamefont
  {Kosteleck\'y}}\ and\ \bibinfo {author} {\bibfnamefont {M.}~\bibnamefont
  {Mewes}},\ }\href {\doibase 10.1016/j.physletb.2016.04.040} {\bibfield
  {journal} {\bibinfo  {journal} {Phys. Lett. B}\ }\textbf {\bibinfo {volume}
  {757}},\ \bibinfo {pages} {510} (\bibinfo {year} {2016})},\ \Eprint
  {http://arxiv.org/abs/1602.04782} {arXiv:1602.04782 [gr-qc]} \BibitemShut
  {NoStop}%
\bibitem [{\citenamefont {Samajdar}\ and\ \citenamefont
  {Arun}(2017)}]{Samajdar:2017mka}%
  \BibitemOpen
  \bibfield  {author} {\bibinfo {author} {\bibfnamefont {A.}~\bibnamefont
  {Samajdar}}\ and\ \bibinfo {author} {\bibfnamefont {K.~G.}\ \bibnamefont
  {Arun}},\ }\href {\doibase 10.1103/PhysRevD.96.104027} {\bibfield  {journal}
  {\bibinfo  {journal} {Phys. Rev. D}\ }\textbf {\bibinfo {volume} {96}},\
  \bibinfo {pages} {104027} (\bibinfo {year} {2017})},\ \Eprint
  {http://arxiv.org/abs/1708.00671} {arXiv:1708.00671 [gr-qc]} \BibitemShut
  {NoStop}%
\bibitem [{\citenamefont {Talbot}\ and\ \citenamefont
  {Thrane}(2018)}]{Talbot:2018cva}%
  \BibitemOpen
  \bibfield  {author} {\bibinfo {author} {\bibfnamefont {C.}~\bibnamefont
  {Talbot}}\ and\ \bibinfo {author} {\bibfnamefont {E.}~\bibnamefont
  {Thrane}},\ }\href {\doibase 10.3847/1538-4357/aab34c} {\bibfield  {journal}
  {\bibinfo  {journal} {Astrophys. J.}\ }\textbf {\bibinfo {volume} {856}},\
  \bibinfo {pages} {173} (\bibinfo {year} {2018})},\ \Eprint
  {http://arxiv.org/abs/1801.02699} {arXiv:1801.02699 [astro-ph.HE]}
  \BibitemShut {NoStop}%
\bibitem [{\citenamefont {Abbott}\ \emph
  {et~al.}(2021{\natexlab{c}})\citenamefont {Abbott} \emph
  {et~al.}}]{GWTC2-pop}%
  \BibitemOpen
  \bibfield  {author} {\bibinfo {author} {\bibfnamefont {R.}~\bibnamefont
  {Abbott}} \emph {et~al.} (\bibinfo {collaboration} {LIGO Scientific,
  Virgo}),\ }\href {\doibase 10.3847/2041-8213/abe949} {\bibfield  {journal}
  {\bibinfo  {journal} {Astrophys. J. Lett.}\ }\textbf {\bibinfo {volume}
  {913}},\ \bibinfo {pages} {L7} (\bibinfo {year} {2021}{\natexlab{c}})},\
  \Eprint {http://arxiv.org/abs/2010.14533} {arXiv:2010.14533 [astro-ph.HE]}
  \BibitemShut {NoStop}%
\bibitem [{\citenamefont {Abbott}\ \emph
  {et~al.}(2023{\natexlab{b}})\citenamefont {Abbott} \emph
  {et~al.}}]{GWTC3-pop}%
  \BibitemOpen
  \bibfield  {author} {\bibinfo {author} {\bibfnamefont {R.}~\bibnamefont
  {Abbott}} \emph {et~al.} (\bibinfo {collaboration} {KAGRA, VIRGO, LIGO
  Scientific}),\ }\href {\doibase 10.1103/PhysRevX.13.011048} {\bibfield
  {journal} {\bibinfo  {journal} {Phys. Rev. X}\ }\textbf {\bibinfo {volume}
  {13}},\ \bibinfo {pages} {011048} (\bibinfo {year} {2023}{\natexlab{b}})},\
  \Eprint {http://arxiv.org/abs/2111.03634} {arXiv:2111.03634 [astro-ph.HE]}
  \BibitemShut {NoStop}%
\bibitem [{\citenamefont {Fishbach}\ and\ \citenamefont
  {Holz}(2020)}]{Fishbach:2019bbm}%
  \BibitemOpen
  \bibfield  {author} {\bibinfo {author} {\bibfnamefont {M.}~\bibnamefont
  {Fishbach}}\ and\ \bibinfo {author} {\bibfnamefont {D.~E.}\ \bibnamefont
  {Holz}},\ }\href {\doibase 10.3847/2041-8213/ab7247} {\bibfield  {journal}
  {\bibinfo  {journal} {Astrophys. J. Lett.}\ }\textbf {\bibinfo {volume}
  {891}},\ \bibinfo {pages} {L27} (\bibinfo {year} {2020})},\ \Eprint
  {http://arxiv.org/abs/1905.12669} {arXiv:1905.12669 [astro-ph.HE]}
  \BibitemShut {NoStop}%
\bibitem [{\citenamefont {Wysocki}\ \emph {et~al.}(2019)\citenamefont
  {Wysocki}, \citenamefont {Lange},\ and\ \citenamefont
  {O'Shaughnessy}}]{Wysocki:2018mpo}%
  \BibitemOpen
  \bibfield  {author} {\bibinfo {author} {\bibfnamefont {D.}~\bibnamefont
  {Wysocki}}, \bibinfo {author} {\bibfnamefont {J.}~\bibnamefont {Lange}}, \
  and\ \bibinfo {author} {\bibfnamefont {R.}~\bibnamefont {O'Shaughnessy}},\
  }\href {\doibase 10.1103/PhysRevD.100.043012} {\bibfield  {journal} {\bibinfo
   {journal} {Phys. Rev. D}\ }\textbf {\bibinfo {volume} {100}},\ \bibinfo
  {pages} {043012} (\bibinfo {year} {2019})},\ \Eprint
  {http://arxiv.org/abs/1805.06442} {arXiv:1805.06442 [gr-qc]} \BibitemShut
  {NoStop}%
\bibitem [{\citenamefont {Madau}\ and\ \citenamefont
  {Dickinson}(2014)}]{Madau:2014bja}%
  \BibitemOpen
  \bibfield  {author} {\bibinfo {author} {\bibfnamefont {P.}~\bibnamefont
  {Madau}}\ and\ \bibinfo {author} {\bibfnamefont {M.}~\bibnamefont
  {Dickinson}},\ }\href {\doibase 10.1146/annurev-astro-081811-125615}
  {\bibfield  {journal} {\bibinfo  {journal} {Ann. Rev. Astron. Astrophys.}\
  }\textbf {\bibinfo {volume} {52}},\ \bibinfo {pages} {415} (\bibinfo {year}
  {2014})},\ \Eprint {http://arxiv.org/abs/1403.0007} {arXiv:1403.0007
  [astro-ph.CO]} \BibitemShut {NoStop}%
\bibitem [{\citenamefont {Abbott}\ \emph
  {et~al.}(2021{\natexlab{d}})\citenamefont {Abbott} \emph
  {et~al.}}]{NSBH-LVK}%
  \BibitemOpen
  \bibfield  {author} {\bibinfo {author} {\bibfnamefont {R.}~\bibnamefont
  {Abbott}} \emph {et~al.} (\bibinfo {collaboration} {LIGO Scientific, KAGRA,
  VIRGO}),\ }\href {\doibase 10.3847/2041-8213/ac082e} {\bibfield  {journal}
  {\bibinfo  {journal} {Astrophys. J. Lett.}\ }\textbf {\bibinfo {volume}
  {915}},\ \bibinfo {pages} {L5} (\bibinfo {year} {2021}{\natexlab{d}})},\
  \Eprint {http://arxiv.org/abs/2106.15163} {arXiv:2106.15163 [astro-ph.HE]}
  \BibitemShut {NoStop}%
\end{thebibliography}%

\clearpage

\onecolumngrid
\appendix*

\section{Supplemental Materials}
In this Supplement, we briefly describe the three types of compact binary populations considered in the paper. The reader must refer to Sec.~III of Ref.~\cite{Gupta2023mpsac} for a more detailed description. We also provide the Bayesian framework for mapping the bounds on multipole deformation parameters $\{\delta \mu_l, \, \delta \epsilon_l\}$ to the PN phasing deformation parameters $\delta \hat{\phi}_b$.

\subsection{Compact binary populations}
\subsubsection{Binary black holes}\label{sec:BBHpop}
The primary black hole masses are drawn from the {\tt Power Law + Peak} mass model~\cite{Talbot:2018cva,GWTC2-pop} with the following values of model parameters: $\alpha=-3.4$, $m_{\rm min}=5\, M_\odot$, $m_{\rm max}=87\, M_\odot$, $\lambda=0.04$, $\mu_{m}=34\, M_\odot$, $\sigma_{m}=3.6\, M_\odot$, and $\delta_{m}=4.8\, M_\odot$~\cite{GWTC3-pop} [See Eq.~(B3) in Appendix B of Ref.~\cite{GWTC3-pop}]. The mass ratio follows a power-law distribution~\cite{Fishbach:2019bbm} with power law index of 1.1~\cite{GWTC3-pop}, and respect the condition $m_{\rm min}=5\, M_\odot$ [See Eq.~(B7) in Appendix B of Ref.~\cite{GWTC3-pop}]. The aligned spins components of the binary ($\chi_{1z}$, $\chi_{2z}$) are drawn from a Beta distribution~\cite{Wysocki:2018mpo} with $\alpha_{\chi}=2$, $\beta_{\chi}=5$~\cite{GWTC3-pop} [See Eq.~(10) of Ref.~\cite{Wysocki:2018mpo}]. The merger rate of the binary black hole population is assumed to follow the Madau-Dickinson star formation rate~\cite{Madau:2014bja} with a local merger rate density of 24 $\rm{Gpc^{-3}\, yr^{-1}}$~\cite{GWTC2-pop}.

\subsubsection{Intermediate mass binary black holes}\label{sec:IMBBHpop}
The masses of the intermediate mass binary black hole population are drawn from a power law distribution with a power-law index of $-2.5$. The lightest and heaviest masses in the distribution are opted to be $m_{\rm min}=100\, M_\odot$ and $m_{\rm max}=1000\, M_\odot$, respectively. The spin components along the orbital angular momentum follow a uniform distribution between [$ -0.9,\, 0.9$]. The merger rate is chosen to follow the Madau-Dickinson star formation rate~\cite{Madau:2014bja} with a local merger rate density of 1 $\rm{Gpc^{-3}\, yr^{-1}}$.

\subsubsection{Neutron star-black hole binaries}\label{sec:NSBHpop}
The black hole masses are drawn from the {\tt Power Law + Peak} mass model~\cite{Talbot:2018cva,GWTC2-pop}, same as the primary mass of the binary black hole population. The masses of the neutron star follow a uniform distribution between $[1, 2.2] \, M_{\odot}$. The aligned spin components of black hole
are assumed to follow a normal distribution with mean 0 and standard deviation of 0.2. 
The aligned spin components of the neutron star are uniformly drawn between $[-0.1, 0.1]$. The merger rate follow the Madau-Dickinson star formation rate~\cite{Madau:2014bja} with a local merger rate density of 45 $\rm{Gpc^{-3}\, yr^{-1}}$~\cite{GWTC3-pop, NSBH-LVK}.

\subsection{Mapping the multipole deformation bounds to the PN phase deformation parameters}\label{sec:mapping}
In the Bayesian framework, measuring the PN phase deformation parameter $\delta \hat{\phi}_b$ amounts to obtaining the posterior probability density function $P(\delta \hat{\phi}_b | d, \mathcal{H})$, where $d$ denotes the detector data and $\mathcal{H}$ denotes the model. Using Bayes' theorem
\begin{equation}\label{eq:Bayes}
    P(\delta \hat{\phi}_b | d, \mathcal{H})  =  \frac{P(\delta \hat{\phi}_b | \mathcal{H}) \, P(d | \delta \hat{\phi}_b, \mathcal{H} )}{P(d | \mathcal{H})},
\end{equation}
where, $P (\delta \hat{\phi}_b | \mathcal{H})$ is the prior probability density function, $P(d | \delta \hat{\phi}_b, \mathcal{H} )$ is the likelihood function, and  $P(d | \mathcal{H})$ [with $P(d | \mathcal{H})= \int d(\delta \hat{\phi}_b) P (\delta \hat{\phi}_b | \mathcal{H}) \, P(d | \delta \hat{\phi}_b, \mathcal{H} )$] is the evidence.

In the parametrized multipolar approach~\cite{Kastha:2018bcr,Kastha:2019} the different gravitational wave phasing coefficients $\phi_{b}$ (also the amplitude~\cite{Mahapatra:2023ydi}) are functions of $\mu_{l}$, $\epsilon_{l}$ along with the other binary's intrinsic parameters. Therefore, different $\delta \hat{\phi}_b$ are also function of $\{\delta \mu_l, \, \delta \epsilon_l\}$ and the intrinsic parameters of binary such as $\nu, \, \chi_{1z},  \, {\rm and} \, \chi_{2z}$. Here, we are interested in computing the posterior probability distribution function of $\delta \hat{\phi}_b$, ${\widetilde P}(\delta \hat{\phi}_b |d, \, \mathcal{H})$, assuming a uniform prior on $\delta \hat{\phi}_b$ [i.e., $P(\delta \hat{\phi}_b | \mathcal{H})={\widetilde \Pi}(\delta \hat{\phi}_b | \mathcal{H})$] from the posterior distributions of $\{\delta \mu_l, \, \delta \epsilon_l\}$ and intrinsic binary parameters. The posterior probability function ${\widetilde P}(\delta \hat{\phi}_b |d, \, \mathcal{H})$ can be obtained by replacing $P(\delta \hat{\phi}_b | \mathcal{H})={\widetilde \Pi}(\delta \hat{\phi}_b | \mathcal{H})$ in Eq.~(\ref{eq:Bayes}) and can be expressed as follows
\begin{align} \label{eq:Mapping-part1}
\widetilde{P}(\delta \hat{\phi}_{b} | d, \mathcal{H}) & =  \frac{{\widetilde \Pi}(\delta \hat{\phi}_{b} | \mathcal{H}) \, P(d | \delta \hat{\phi}_{b} , \mathcal{H})}{{\widetilde P}(d | \mathcal{H})} \nonumber\\
& = \frac{{\widetilde \Pi}(\delta \hat{\phi}_{b} | \mathcal{H}) \, \int d \vec{\lambda}_{I} \, d\vec{\lambda}_{T} \, P(d| \vec{\lambda}_{I}, \vec{\lambda}_{T}, \mathcal{H}) \, P(\vec{\lambda}_{I}, \vec{\lambda}_{T} | \delta \hat{\phi}_{b}, \mathcal{H})}{{\widetilde P}(d | \mathcal{H})} \nonumber\\
& = \frac{{\widetilde \Pi}(\delta \hat{\phi}_{b} | \mathcal{H})}{{\widetilde P}(d | \mathcal{H})} \, \times \, \int d \vec{\lambda}_{I} \, d\vec{\lambda}_{T} P(d| \vec{\lambda}_{I}, \vec{\lambda}_{T}, \mathcal{H}) \, \underbrace{\frac{{\widetilde \Pi}( \vec{\lambda}_{I}, \vec{\lambda}_{T} | \mathcal{H}) \, P(\delta \hat{\phi}_{b} |  \vec{\lambda}_{I}, \vec{\lambda}_{T}, \mathcal{H})}{\Pi (\delta \hat{\phi}_{b}| \mathcal{H})}}_{P(\vec{\lambda}_{I}, \vec{\lambda}_{T} | \delta \hat{\phi}_{b}, \mathcal{H})} \, ,
\end{align}
where, the intrinsic binary parameters are represented by $\vec{\lambda}_{I} \in \{\nu, \, \chi_{1z}, \, \chi_{2z}  \}$, the mulitpolar coefficients are denoted by $\vec{\lambda}_{T} \in \{\delta \mu_{l}, \, \delta \epsilon_{l}\}$. ${\widetilde P}(d|\mathcal{H})$ is the evidence for the uniform prior on $\delta \hat{\phi}_{b}$, $P(d| \vec{\lambda}_{I}, \vec{\lambda}_{T}, \mathcal{H})$ is the likelihood function of the gravitational wave data given the parameters $\{ \vec{\lambda}_{I}, \, \vec{\lambda}_{T}\}$, $P(\delta \hat {\phi_b} \, | \, \vec{\lambda}_{I}, \, \vec{\lambda}_{T}, \, \mathcal{H})$ takes care of the coordinate transformation between $\{ \vec{\lambda}_{I}, \,  \vec{\lambda}_{T}\}$ and $\delta \hat{\phi}_{b}$, and $\Pi (\delta \hat{\phi}_{b}| \mathcal{H})$ is given by
\begin{equation}\label{eq:old-prior}
    \Pi (\delta \hat{\phi}_{b} | \mathcal{H}) \equiv \int d \vec{\lambda}_{I} \, d\vec{\lambda}_{T} \, {\widetilde \Pi}( \vec{\lambda}_{I}, \vec{\lambda}_{T} | \mathcal{H}) \, P(\delta \hat{\phi}_{b} |  \vec{\lambda}_{I}, \vec{\lambda}_{T}, \mathcal{H})\, .
\end{equation}
Therefore, $\Pi(\delta \hat{\phi}_{b} \, | \, \mathcal{H} )$ is simply the distribution of $\delta \hat{\phi}_{b}$ derived from the uniform prior on $\vec{\lambda}_{I}$ and $\vec{\lambda}_{T}$ and the relation between $\delta \hat{\phi}_{b}$ and \{$\vec{\lambda}_{I}$, $\vec{\lambda}_{T}$\}. In Eq.~(\ref{eq:Mapping-part1}), we have used Bayes' theorem which can be further simplified as follows
\begin{align} \label{eq:Mapping}
\widetilde{P}(\delta \hat{\phi}_{b} | d, \mathcal{H}) & = {\widetilde \Pi}(\delta \hat{\phi}_{b} | \mathcal{H}) \, \times \frac{{\widetilde P}_{IT}(d|\mathcal{H})}{{\widetilde P}(d|\mathcal{H})} \times \, \int d \vec{\lambda}_{I} \, d\vec{\lambda}_{T} \bigg[ \underbrace{\frac{{\widetilde \Pi}( \vec{\lambda}_{I}, \vec{\lambda}_{T} | \mathcal{H}) \, P(d| \vec{\lambda}_{I}, \vec{\lambda}_{T},  \mathcal{H})}{{\widetilde P}_{IT}(d | \mathcal{H})}}_{\widetilde{P}(\vec{\lambda}_{I}, \vec{\lambda}_{T} | d, \mathcal{H})} \, \times \, \frac{P(\delta \hat{\phi}_{b} | \vec{\lambda}_{I}, \vec{\lambda}_{T}, \mathcal{H})}{\Pi (\delta \hat{\phi}_{b}|\mathcal{H})} \bigg] \nonumber \\
& = \bigg[ \int d \vec{\lambda}_{I} \, d\vec{\lambda}_{T} P(\delta \hat{\phi}_{b}|\vec{\lambda}_{I}, \, \vec{\lambda}_{T}, \,  \mathcal{H})\;{\widetilde P}( \vec{\lambda}_{I}, \, \vec{\lambda}_{T} |d, \, \, \mathcal{H})\;\bigg] \times \frac{{\widetilde \Pi}(\delta \hat{\phi}_{b} | \mathcal{H}) }{\Pi (\delta \hat{\phi}_{b}|\mathcal{H})} \times \frac{{\widetilde P}_{IT}(d|\mathcal{H})}{{\widetilde P}(d|\mathcal{H})},
\end{align}
where, ${\widetilde P}(\vec{\lambda}_{I}, \, \vec{\lambda}_{T}\, | \, d, \, \mathcal{H})$ is the posterior probability density of $\{ \vec{\lambda}_{I}, \,  \vec{\lambda}_{T}\}$ and ${\widetilde P}_{IT}(d|\mathcal{H})$ [with ${\widetilde P}_{IT}(d|\mathcal{H}) = \int d \vec{\lambda}_{I} \, d \vec{\lambda}_{T} \, {\widetilde \Pi}( \vec{\lambda}_{I}, \vec{\lambda}_{T} | \mathcal{H}) \, P(d| \vec{\lambda}_{I}, \vec{\lambda}_{T},  \mathcal{H})$] is the corresponding evidence derived assuming a uniform prior on $\{ \vec{\lambda}_{I}, \,  \vec{\lambda}_{T}\}$. The numerical factor $\frac{{\widetilde P}_{IT}(d|\mathcal{H})}{{\widetilde P}(d|\mathcal{H})}$ is an overall normalization constant in the above equation and can be ignored if one is only interested in estimating $\widetilde{P}(\delta \hat{\phi}_{b} | d, \mathcal{H})$. The different steps for the derivation of the above equation are followed from Ref.~\cite{Mahapatra:2023hqq} (See Sec.~2 of the Supplemental Material in Ref.~\cite{Mahapatra:2023hqq}).

As mentioned earlier, $\delta \hat{\phi}_{b}$ is a unique function of $\{ \vec{\lambda}_{I}, \, \vec{\lambda}_{T}\} $, say $f_{b}(\vec{\lambda}_{I}, \, \vec{\lambda}_{T})$~\cite{Kastha:2018bcr,Kastha:2019}; hence given a value of $\{ \vec{\lambda}_{I}, \, \vec{\lambda}_{T}\}$, $\delta \hat{\phi}_{b}$ is particularly determined. Therefore, $P(\delta \hat {\phi_b} \, | \, \vec{\lambda}_{I}, \, \vec{\lambda}_{T}, \, \mathcal{H})$ can simply be represented by a delta function,
\begin{equation}
    P(\delta \hat {\phi_b} \, | \, \vec{\lambda}_{I}, \, \vec{\lambda}_{T}, \, \mathcal{H}) = \delta (\delta \hat {\phi_b}- f_{b}(\vec{\lambda}_{I}, \, \vec{\lambda}_{T})) \, .
\end{equation}
In practice, to compute $\widetilde{P}(\delta \hat{\phi}_{b} | d, \mathcal{H})$ one needs to take the posterior samples of $\{ \vec{\lambda}_{I}, \,  \vec{\lambda}_{T}\}$ and estimate $\delta \hat {\phi_b}$ for each sample through the functions $f_{b}(\vec{\lambda}_{I}, \, \vec{\lambda}_{T})$; and then reweight these samples by the probability $\frac{{\widetilde \Pi}(\delta \hat {\phi_b} | \mathcal{H}) }{\Pi (\delta \hat {\phi_b}|\mathcal{H})}$.

\end{document}